\documentclass[%
 jor,
 prl,
 amsmath,amssymb,
twocolumn,
]{revtex4-2}

\usepackage{graphicx}
\usepackage{dcolumn}
\usepackage{bm}
\usepackage{bbold}
\usepackage{float}

\usepackage{color,soul}

\renewcommand{\vec}[1]{\bm{\mathbf{#1}}}

\begin{document}

\title{Magnetic Phase Diagram of YbRh\textsubscript{2}Si\textsubscript{2}:\\
the Influence of Hyperfine Interactions
}

\author{J.\ Knapp}
\affiliation{
Department of Physics, Royal Holloway University of London, TW20 0EX, Egham, UK.
}
\email{\mbox{jan.knapp@rhul.ac.uk, l.v.levitin@rhul.ac.uk,} j.saunders@rhul.ac.uk}
\author{L.\ V.\ Levitin}
\affiliation{ 
Department of Physics, Royal Holloway University of London, TW20 0EX, Egham, UK.
}
\email{l.v.levitin@rhul.ac.uk}
\author{J.\ Ny\'eki}
\affiliation{ 
Department of Physics, Royal Holloway University of London, TW20 0EX, Egham, UK.
}
\author{B.\ Cowan}
\affiliation{
Department of Physics, Royal Holloway University of London, TW20 0EX, Egham, UK.
}
\author{J.\ Saunders}
\affiliation{
Department of Physics, Royal Holloway University of London, TW20 0EX, Egham, UK.
}
\email{j.saunders@rhul.ac.uk.}
\author{M.\ Brando}%
\affiliation{ 
Max Planck Institute for Chemical Physics of Solids, N\"othnitzer Stra\ss{}e 40, 01187 Dresden, Germany.
}
\author{C.\ Geibel}%
\affiliation{ 
Max Planck Institute for Chemical Physics of Solids, N\"othnitzer Stra\ss{}e 40, 01187 Dresden, Germany.
}

\author{K.\ Kliemt}%
\affiliation{ 
Physikalisches Institut, Max-von-Laue-Stra\ss{}e 1, 60438 Frankfurt am Main, Germany.
}
\author{C.\ Krellner}%
\affiliation{ 
Physikalisches Institut, Max-von-Laue-Stra\ss{}e 1, 60438 Frankfurt am Main, Germany.
}

\date{\today}

\begin{abstract}
We report the determination of the magnetic phase diagram of the heavy fermion metal YbRh\textsubscript{2}Si\textsubscript{2} in magnetic fields up to 70\,mT applied perpendicular to the crystallographic c-axis.
By a combination of heat capacity, magneto-caloric, and magneto-resistance measurements we map two antiferromagnetic phases: the electronic AFM1 below 70\,mK and electro-nuclear AFM2 below 1.5\,mK.
The measurements extend into the microkelvin regime to explore the quantum phase transitions in this system.
We demonstrate how the hyperfine interaction significantly modifies the phase diagram and the putative field-tuned quantum critical point.
The determination of the rich magnetic properties of YbRh\textsubscript{2}Si\textsubscript{2} is essential to understanding the interplay of the two magnetic orders and superconductivity in this compound.
\end{abstract}


\maketitle


The heavy fermion metal YbRh\textsubscript{2}Si\textsubscript{2} has been extensively studied for many years as an example of a system that can be tuned towards quantum criticality, by chemical pressure or by magnetic field~\cite{Gegenwart2002,Custers2003,Pfau2012,Gegenwart2007,Gegenwart2008,Friedemann2009,Knebel2006}.
Often, superconductivity emerges in heavy fermion metals when they are tuned to the quantum critical point (QCP);
the effect of pressure on multiple Ce-based compounds being well established~\cite{Weng2016}.
Following the first Yb-based superconducting heavy fermion $\beta$-YbAlB\textsubscript{4}~\cite{Nakatsuji2008},
more recently the discovery of superconductivity in YbRh\textsubscript{2}Si\textsubscript{2}~\cite{Schuberth2016,Nguyen2021,Levitin2025} makes it only the second example of an Yb-based superconductor.
In a material where superconductivity emerges out of magnetically ordered states, like YbRh\textsubscript{2}Si\textsubscript{2}, establishing the magnetic phase diagram is essential.

\begin{figure}[!b]
    \centering
    \includegraphics{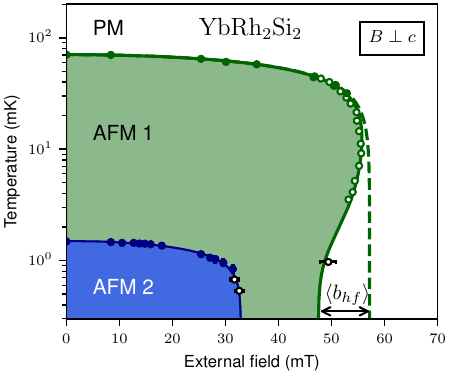}
    \caption{
    Phase diagram of the two antiferromagnetic orders in YbRh\textsubscript{2}Si\textsubscript{2}.
    The critical field $B_N(T)$ of the AFM1 phase (solid green line) shows a non-monotonic temperature dependence.
    We understand this $T\rightarrow 0$ backturn from the asymptote $B_N^*$ (dashed green line) as an effect of the hyperfine field $\langle b_{\text{hf}}\rangle$, which adds to the external field according to Eq.~\eqref{eq:B_eff}.
    The critical field $B_A(T)$ of the electro-nuclear AFM2 phase shows a regular behavior.
    The heat capacity peaks are shown by filled circles, magneto-caloric sweep signature by open black circles, and magneto-resistance by open green circles.
    Calorimetry and transport studies were done on two different samples.}
    \label{fig:Phase_diagram}
\end{figure}

Here we report a determination of the phase diagram extending into the \textmu K regime, Fig.~\ref{fig:Phase_diagram}.
It features two antiferromagnetic phases: electronic AFM1~\cite{Gegenwart2002} and electro-nuclear AFM2~\cite{Knapp2023},
each terminating at a field-tuned quantum phase transition.
Previously the critical field of AFM1 had only been mapped down to 20\,mK~\cite{Gegenwart2002}.
We study tetragonal single crystals of YbRh\textsubscript{2}Si\textsubscript{2} with natural isotopic composition, with magnetic field applied in the ab-plane.
The measurements down to ultra-low temperatures
required novel calorimetry and electrical transport experimental techniques~\cite{Knapp2024,Levitin2025}.
Both the observed backturn of the critical field of AFM1 and the stability of AFM2 are distinct manifestations of the hyperfine interaction of Yb.

YbRh\textsubscript{2}Si\textsubscript{2} features strong anisotropy of the Fermi surface \cite{Friedemann2010b,Kummer2015,Zwicknagl2016,Li2019a,Guettler2021,Knebel2006} and electronic magnetism~\cite{Gegenwart2002}.
The AFM1 phase, suppressed by in-plane field of 57\,mT or an order of magnitude larger field along the c-axis \cite{Gegenwart2002}, features only very small ordered moments, $\mu_{e}\approx0.002\,\mu_{\mathrm{B}}$~\cite{Ishida2003}, developing out of partially Kondo-screened Yb local moments $1.4\,\mu_{\mathrm{B}}$~\cite{Gegenwart2002}, where $\mu_{\mathrm{B}}$ is the Bohr magneton.
Ordering of the small moments along the magnetically-hard c-axis has been proposed~\cite{Hamann2019}, but the nature of this order is not understood, and neither is the putative quantum criticality proposed for the suppression of this order with magnetic field.
Current theories include the \textit{local quantum criticality}~\cite{Si2001,Coleman2005,Si2010,Steglich2014}, see also \cite{Schubert2019,Gegenwart2007,Gegenwart2008,Paschen2020,Schuberth2022}, and theories invoking strong coupling of fermions and spin fluctuations into critical quasiparticles~\cite{Abrahams2012,Abrahams2014,Woelfle2015,Watanabe2010}.

\begin{figure}[!t]
    \centering
    \includegraphics{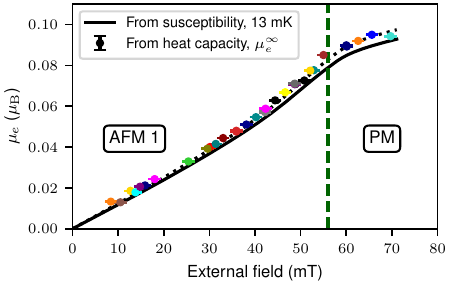}
    \caption{Growth of the polarised electronic magnetic moment $\mu_e$ with in-plane magnetic field
    at 13\,mK, calculated as the integral of magnetic susceptibility~\cite{Steinke2017},
    and determined from the heat capacity using the \textit{Hyperfine Schottky model}~\cite{SM}.
    Scaling the susceptibility data by a factor of 1.05 (dotted line) removes a small systematic discrepancy between the two methods.
    }
    \label{fig:4f_inf_vs_H_dynamic}
\end{figure}

A remarkable property of the AFM1 phase is the strong paramagnetic response to in-plane field $B_{\mathrm{ext}}$, manifested by a rapid growth of a static electronic moment $\mu_e$.
Figure~\ref{fig:4f_inf_vs_H_dynamic} shows excellent agreement of $\mu_e(B_{\mathrm{ext}})$, inferred from the nuclear heat capacity by a method discussed in detail later in this Letter, with the integral of the AC magnetic susceptibility $\chi$ at 13\,mK reported in Ref.~\cite{Steinke2017}.
This agreement identifies the 4f electronic orbital of Yb as the origin of the dominant magnetic effects in YbRh\textsubscript{2}Si\textsubscript{2}.
The measurements give a single Yb ion susceptibility $\chi_{\mathrm{ion}}\equiv \mu_e/B_{\mathrm{ext}} \approx1.2\,\mu_{\mathrm{B}}$/T (at 13\,mK and zero field), equivalent to the dimensionless susceptibility $\chi \approx 0.18$.
Two properties of the $\mu_e(B_{\mathrm{ext}})$ curve, the inflection at 50\,mT and change of slope at 57\,mT reflect two features in $\chi(B_{\mathrm{ext}})$ labeled ``$B_1$'' and ``$B_2$'' in Ref.~\cite{Steinke2017}; the latter feature has been associated with the critical field of AFM1.
Thus established $\mu_e(B_{\mathrm{ext}})$ is the key input for our analysis of the influence of the hyperfine interaction on the AFM1/PM phase boundary.

Recently we identified the phase, here denoted AFM2, as an electro-nuclear spin density wave, the transition to which occurs as a direct result of the hyperfine interaction~\cite{Knapp2023}.
The staggered magnetic moments were found to lie in the ab-plane and to be continuously suppressed from $\mu_e \sim 0.1\,\mu_{\mathrm{B}}$ to zero by in-plane magnetic field.
Ref.~\cite{Knapp2023} also introduces the use of the heat capacity of nuclear spins as a probe of the electronic moments hyperfine-coupled to them, a method we advance here.

\begin{figure}[!b]
    \centering
    \includegraphics{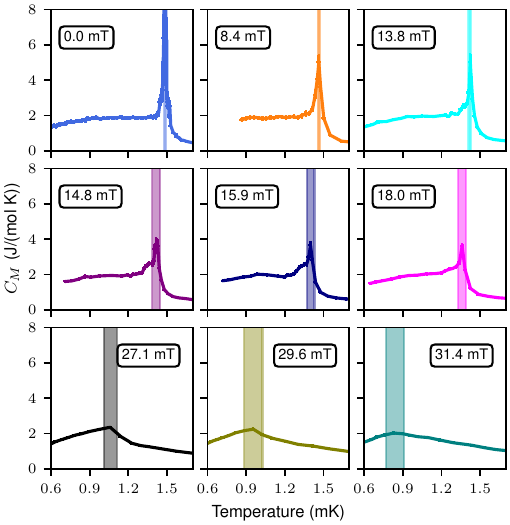}
    \includegraphics{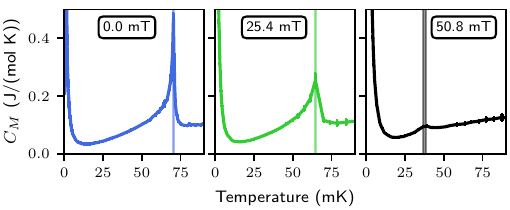}
    \caption{
    Evolution of the $T_A$ (top) and $T_N$ (bottom) transition peaks in the molar heat capacity $C_M$ with magnetic field.
    Estimated transition temperature and its uncertainty are marked by vertical bands.
    Weak effects of measurement preparation conditions on the position and shape of the $T_A$ transition signature are discussed in SM~\cite{SM}.
    }
    \label{fig:Ordering_peaks}
\end{figure}

\begin{figure}[!t]
    \centering
    \includegraphics{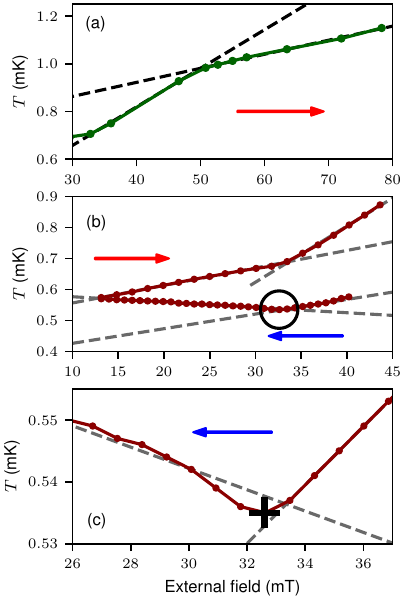}
    \caption{Examples of quasi-adiabatic magneto-caloric sweeps with linear fits above and below the transition.
    Direction of the field change indicated by the arrows.
    (a) Magnetization sweep from AFM1 to PM.
    (b) De-)magnetization sweeps across the AFM1/AFM2 boundary.
        The negative slope is a consequence of parasitic heat deposition into the sample.
    (c) Detail of the de-magnetization sweep in (b) highlighted by the circle.
    A small dip is observed at the AFM1/AFM2 phase transition (in both directions) and here marked by the cross.
    These features are used as transition points in Fig.~\ref{fig:Phase_diagram}.
    }
    \label{fig:MC_sweeps_glob}
\end{figure}

Our primary experimental technique is calorimetry~\cite{Knapp2024}, modified from Ref.~\cite{Knapp2023} to allow adiabatic measurements to higher fields.
The heat capacity peaks at $T_A$ (onset of AFM2) and $T_N$ (onset of AFM1), Fig.~\ref{fig:Ordering_peaks}, give extremely sharp signatures of the transitions in low external field $B_{\mathrm{ext}}$.
In both cases the peaks are suppressed in magnitude with increasing external field.
The cusp at $T_A$ in high fields closely resembles the predictions of the mean-field theory~\cite{Knapp2023}.
The sharpness in zero field may be related to the concurrent transition in the superconducting state~\cite{Levitin2025}.

\begin{figure}[!t]
    \centering
    \includegraphics{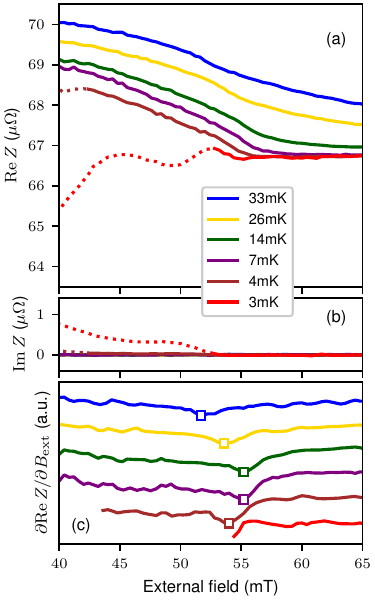}
    \caption{Measurements of magneto-resistance (a) identify the AFM1/PM phase boundary by a minimum in $\partial\mathrm{Re}\,Z/\partial B_{\text{ext}}$~(c).
    (b) At the lowest temperature this procedure is interrupted by the onset of superconductivity,
    revealed by the appearance of $\mathrm{Im}\,Z$ associated with kinetic inductance~\cite{Levitin2025}.
    On these grounds the data where $\mathrm{Im}\,Z > 0.05\,\mu\Omega$ (shown with dotted lines) were excluded from the analysis.
    The superconductivity in YbRh$_2$Si$_2$ is heterogeneous and does not always manifest as $\mathrm{Re}\,Z = 0$~\cite{Levitin2025}.}
    \label{fig:Magnetoresistance}
\end{figure}

When $\partial T_{A/N}/\partial B_{\mathrm{ext}}$ is high, we must use field sweep methods to identify the phase transitions:
the magneto-caloric sweeps, Fig.~\ref{fig:MC_sweeps_glob}, and magneto-resistance, Fig.~\ref{fig:Magnetoresistance}.
The quasi-adiabatic magneto-caloric field sweeps~\cite{Knapp2024} allow us to pin-point the AFM1/PM and AFM2/AFM1 transitions at a temperature below 1\,mK, thus close to the quantum phase transition end-points.
The observed magneto-caloric effect arises from the entropy of Yb nuclear spins,
hyperfine-coupled to $\mu_e$~\cite{SM}, while the electronic entropy is largely quenched.
This is in contrast to strong variation of electronic entropy in the vicinity of phase transitions in materials explored for magnetic refrigeration~\cite{GschneidnerJr2005,Pecharsky2001,Pecharsky1997}.
Despite a parasitic heat deposition into the sample throughout the sweep,
clear features are observed in the response of temperature to the varying field.
Crossing the critical field of AFM1, see Fig.~\ref{fig:MC_sweeps_glob}a,
the change of slope of the adiabatic magneto-caloric sweeps is directly linked~\cite{SM} to the
drop in magnetic susceptibility across the phase boundary~\cite{Steinke2017}.
The AFM1/AFM2 transition is also marked by a change of slope, see Fig.~\ref{fig:MC_sweeps_glob}b,c.
The weaker magneto-caloric effect in the AFM2 phase is a consequence of staggered nuclear polarization
of the spin-density wave, that considerably bounds the nuclear entropy~\cite{Knapp2023}.

The AFM1/PM boundary was also determined between 4 and 45\,mK from magneto-resistance measurements, Fig.~\ref{fig:Magnetoresistance}, exploiting our SQUID-based measurements of the complex sample impedance $Z$
with 10\,n$\Omega$ resolution~\cite{Levitin2025}.
We identify a minimum in $\partial \mathrm{Re}\,Z/\partial B_{\mathrm{ext}}$ as the signature of the phase transition.
This is justified by the coincidence of this magneto-resistance signature with the heat capacity peaks above 32\,mK, see Fig.~\ref{fig:Phase_diagram}.
The magneto-resistance method requires the sample to be in the normal state, so it is interrupted at 3\,mK by the onset of superconductivity, see Fig.~\ref{fig:Magnetoresistance}.

Both field sweep methods exhibit no discernible jumps or hysteresis (see Figs.~\ref{fig:MC_sweeps_glob}, \ref{fig:Magnetoresistance} and Fig.~S5 in SM~\cite{SM}) across the phase boundaries, suggestive of second-order AFM1/PM and AFM1/AFM2 transitions.
This is consistent with the conclusions of earlier work~\cite{Krellner2009,Steinke2017,Knapp2023}.

The AFM2 dome, $B_{A}(T)$, Fig.~\ref{fig:Phase_diagram} is well described by a cubic polynomial fit to $B_{A}^2(T)$, see Eq.~(S12) in SM~\cite{SM}.
This allows us to experimentally determine a critical field of 33\,mT in the $T=0$ limit.
The data are in good agreement with the line of second-order transitions predicted by the mean field model~\cite{Knapp2023}.

The AFM1/PM phase boundary above 20\,mK can also be described by a phenomenological ansatz for $B_{N}^2(T)$ of the same form.
However, at lower temperatures we observe a clear ``backturn'', i.e.\ decrease of the critical field as temperature is lowered, which is fully accounted for by the hyperfine interaction, with the relevant hyperfine field saturating at around 10\,mT at the lowest temperatures.
Consequently, the external critical field at $T=0$ is 47\,mT, rather than 57\,mT conventionally extrapolated from high-temperature measurements~\cite{Gegenwart2002}.

We now explain in detail the procedure to calculate the backturn of the AFM1/PM phase boundary, in terms of known system properties, that is in full agreement with observations, see Fig.~\ref{fig:Phase_diagram}.
The dominant hyperfine interaction in YbRh\textsubscript{2}Si\textsubscript{2} 
involves the 4f electron orbitals of Yb~\cite{Nowik1968,Plessel2003,Knebel2006,Flouquet1975,Flouquet1978}
and arises in the Yb isotopes with non-zero nuclear spin:
$^{171}$Yb ($I=1/2$) and $^{173}$Yb ($I=5/2$) with natural abundances 0.14 and 0.16 respectively.
The hyperfine constant of Yb is particularly strong, $A_{\mathrm{hf}} = 102\pm3\,\mathrm{T}/$\textmu\textsubscript{B}~\cite{Bonville1992,Bonville1991,Bonville1984,Kondo1961}.
In YbRh\textsubscript{2}Si\textsubscript{2} it combines with the particularly strong dependence of 4f~moment on in-plane field $\mu_e(B_{\mathrm{ext}})$ with dramatic effects.

The hyperfine interaction can be expressed as follows:
$U_{\text{hf}} = A_{\text{hf\,}}\vec\mu_e \cdot \vec\mu_n = -\vec B_{\text{hf}} \cdot \vec\mu_n = - \vec b_{\text{hf}} \cdot \vec\mu_e$. Here $\vec B_{\mathrm{hf}}=-A_{\mathrm{hf\,}}\vec\mu_e$ is the hyperfine field
exerted by the magnetic moment $\vec \mu_e$ of the electron on the nuclear magnetic moment $\vec \mu_n$; the reciprocal field 
$\vec b_{\mathrm{hf}}=-A_{\mathrm{hf\,}}\vec\mu_n$
is felt by the electron.
$B_{\mathrm{hf}}$ dominates the nuclear energy level splitting ($B_{\text{hf}} \gg B_{\text{ext}}$) for $^{171}$Yb ($I=1/2$), while in the case of $^{173}$Yb ($I=5/2$) the known quadrupolar interaction~\cite{Knapp2023}, see also Eq.~(S3) in SM~\cite{SM}, must also be taken into account.
The temperature-dependent nuclear polarization is determined self-consistently with $\mu_e(B_{\text{ext}})$, taking into account $b_{\mathrm{hf}}$.
We note that inside AFM1, the staggered magnetization is insignificant and the hyperfine fields are essentially collinear with the external field.

\begin{figure}[t]
    \centering
    \includegraphics[width=0.95\linewidth]{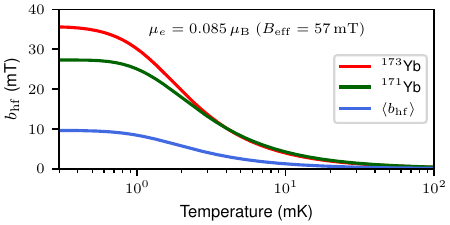}
    \caption{Hyperfine fields on the $^{171}$Yb and $^{173}$Yb sites and the the spatial average $\langle b_{\mathrm{hf}}\rangle$ over all Yb isotopes as a function of temperature for a constant electronic moment $\mu_e = 0.085\,\mu_\mathrm{B}$. This value of $\mu_e$ corresponds to $B^* = 57$\,mT, the critical effective field of AFM1 below 13\,mK.
    The quadrupolar splitting of $^{173}$Yb in the crystalline electric field is taken into account~\cite{Knapp2023}. Details in SM~\cite{SM}.}
    \label{fig:Hyperfine_fields}
\end{figure}

Figure~\ref{fig:Hyperfine_fields} shows an estimate of $b_{\mathrm{hf}}(T)$
for both isotopes at the critical field of AFM1, using the electronic moment $\mu_e = 0.085\,\mu_\mathrm{B}$ measured at this phase boundary at 13\,mK, see Fig.~\ref{fig:4f_inf_vs_H_dynamic}.
We find that the backturn is compensated if we add the average $\langle b_{\mathrm{hf}} \rangle$ over the Yb isotopes to $B_N$.
Thus, we propose that the growth of paramagnetic moments $\mu_e$ and the suppression of AFM1 order is driven by the \emph{effective field}
\begin{equation}
    B^* = B_{\text{ext}} + \langle b_{\mathrm{hf}} \rangle.
\label{eq:B_eff}
\end{equation}
In terms of this effective field the AFM1/PM phase boundary
$B_N^\ast(T) = B_N(T) + \big\langle b_{\mathrm{hf}}\big(T,\mu_e(B_N^\ast(T))\big)\big\rangle$
is well described by a cubic polynomial fit to $B_{N}^{*2}(T)$ over the full temperature range (See Eq.~(S11) in SM~\cite{SM}).
In the temperature-effective field plane this phase boundary approaches the quantum phase transition vertically, with no backturn, see Fig.~S1 in SM~\cite{SM}.

The field dependence of electronic moments $\mu_e(B_{\text{ext}})$ at 13\,mK, shown in Fig.~\ref{fig:4f_inf_vs_H_dynamic}, is inferred  from the heat capacity measured between 1.7 and 13\,mK, taking into account the gradual increase $\mu_e$ on further cooling in a fixed external field driven by $\langle b_{\text{hf}}(T) \rangle$.
This improved \emph{hyperfine Schottky model} resolves the discrepancy between the data and the Schottky model with temperature-independent $\mu_e(B_{\text{ext}})$ observed in Ref.~\cite{Knapp2023}.
We emphasize that $\mu_e$ obtained from this method is in good agreement with the integral of the magnetic  susceptibility~\cite{Steinke2017}, also plotted in Fig.~\ref{fig:4f_inf_vs_H_dynamic}.
From these data we derive $\mu_e(B^\ast)$, assuming it to be temperature independent, and self-consistently calculate $b_{\mathrm{hf}}(B_{\text{ext}},T)$, feeding into the determination of $B_N^*(T)$ from the measured phase transition signatures~\cite{SM,Note1}.%
\footnotetext[1]{The measured susceptibility leads to the enhancement of the magnetic induction $B=(1+\chi)B_{\mathrm{ext}}$ inside YbRh\textsubscript{2}Si\textsubscript{2} over the external field $B_{\mathrm{ext}}$ by about 20\%.
This effect should be incorporated in future theoretical models of the magnetism in YbRh\textsubscript{2}Si\textsubscript{2}.}

Investigating the critical behavior near a QCP~\cite{Continentino1998,Chandra2020,Friedemann2009} can provide an important insight into the nature of the quantum criticality.
We analyze the AFM1/PM transition points as $T = \big(B_N^\ast(T=0) - B_N^\ast(T) \big)^\varepsilon$ and determine the critical exponent $\varepsilon\approx 0.5$,
a similar exercise for the AFM1/AFM2 boundary yields
$\varepsilon\approx0.3$~\cite{SM}.
These values should be treated with caution because of the limited temperature range 
and potential effects of disorder.

A detailed study with $B_{\text{ext}}\parallel c$, requiring field-swept measurements in a new higher-field set-up, is beyond the scope of this work.
A preliminary phase diagram for this field orientation~\cite{Levitin2025}
is significantly different from $B_{\text{ext}}\perp c$,
with evidence for a PM/AFM1/AFM2 polycritical point,
resembling Yb$($Rh$_{0.82}$Co$_{0.18})_2$Si$_2$~\cite{Hamann2019}. The relationship between the magnetic structure of the electro-nuclear order of the AFM2 phase in YbRh\textsubscript{2}Si\textsubscript{2} and the eponymous phase in Ref.~\cite{Hamann2019} remains an open question.

Our results strongly motivate future experiments on isotopically enriched samples using the techniques we have developed.
In the absence of active nuclear spins, for example in \textsuperscript{174}YbRh\textsubscript{2}Si\textsubscript{2}~\cite{Knebel2006}, we predict the external critical field of the AFM1 phase to coincide with the $B_N^*(T)$ line, simplifying the determination of the exponent $\varepsilon$, and the AFM2 phase to be absent.
On the other hand, we estimate enhanced backturn in \textsuperscript{171}YbRh\textsubscript{2}Si\textsubscript{2} and \textsuperscript{173}YbRh\textsubscript{2}Si\textsubscript{2},
where the hyperfine fields saturate in the $T \to 0$ limit at 27\,mT and 35\,mT respectively, see Fig.~\ref{fig:Hyperfine_fields}.
Mean field theory~\cite{Knapp2023} predicts an enhancement in $T_{A}$ by a factor of 3 in zero magnetic field.
Moreover, in these materials the nuclear spins form a regular lattice, in contrast to natural YbRh\textsubscript{2}Si\textsubscript{2}, which is likely to have significant consequences on the AFM2 phase.

The theoretical framework for treating the impact of hyperfine interactions on quantum criticality has recently been reported~\cite{Eisenlohr2021}.
Experimentally, hyperfine effects have been investigated in the transverse-field Ising ferromagnet LiHoF\textsubscript{4}~\cite{Bitko1996,Libersky2021}, where findings include a shift in quantum critical field and softening of collective electro-nuclear modes near the QCP.
In the compound YbCu\textsubscript{4.6}Au\textsubscript{0.4} it was hypothesized that the quantum critical fluctuations do not originate from purely electronic states but from entangled electro-nuclear states, due to the low Kondo and RKKY energy scales~\cite{banda2023}.
Diverse influences of hyperfine interactions on superconductivity and antiferroquadrupolar order have also been found in PrOs\textsubscript{4}Sb\textsubscript{12}~\cite{Bangma2023}. The present results open the door to the study of the influence of hyperfine interactions on putative heavy fermion quantum criticality in YbRh\textsubscript{2}Si\textsubscript{2}.

In conclusion, we report diverse hyperfine effects in YbRh\textsubscript{2}Si\textsubscript{2}, which has been referred to as a ``canonical heavy fermion'' material ~\cite{Schuberth2016}.
One striking finding is that, despite the locality of the hyperfine fields $b_{\mathrm{hf}}$ experienced by the 4f electronic shell on the $^{171}$Yb and $^{173}$Yb sites, the entire electron system responds to the average $\langle b_{\mathrm{hf}}\rangle$.
This offers fresh insight insight into the nature of the strongly correlated heavy fermion state. 
The present results demonstrate YbRh\textsubscript{2}Si\textsubscript{2} as a unique laboratory of heavy fermion physics, featuring interplay between magnetism and superconductivity, and cooperative electro-nuclear phenomena, with isotopic composition a key tuning parameter in future work.

\begin{acknowledgments}
This work was supported by the European Microkelvin Platform, by the European Union's Horizon 2020 Research and Innovation programme under grant agreement no. 824109. It was also supported by the Deutsche Forschungsgemeinschaft (DFG, German Research Foundation) through grants Nos. BR 4110/1-1, KR3831/4-1, and via the TRR 288, (422213477, project A03).
\end{acknowledgments}

\onecolumngrid

\renewcommand{\thefigure}{S\arabic{figure}}
\renewcommand{\thetable}{S\arabic{table}}
\renewcommand{\theequation}{S\arabic{equation}}
\setcounter{figure}{0}
\setcounter{table}{0}
\setcounter{equation}{0}

\vskip2em

\centerline{\large\textbf{Supplemental Material}}
\vskip1em

\twocolumngrid 

\section{Calculation of hyperfine fields}

At the center of the modeling procedures discussed in the following sections is the hyperfine interaction.
We choose the coordinate system with $x$ axis aligned
with the direction of the in-plane magnetic field
and $z$ axis pointing along the c axis.
Then, restricting our considerations to the AFM1 and PM phases
and ignoring the small staggered magnetization of AFM1,
all relevant vectors are aligned with the $x$ axis.
For a given temperature $T$ and 
field of the reverse action $B_{\mathrm{hf}} = -A_{\mathrm{hf}} \mu_e$ we calculate the $x$ components of the hyperfine fields
\begin{equation}
    b_{\mathrm{hf}}^{171/173} = -A_{\mathrm{hf}}\mu_n = -A_{\mathrm{hf}}g\mu_NI_x,
\label{eq:bHF}
\end{equation}
where the static nuclear polarization
\begin{equation}
    I_x = \sum_i p_i \langle\psi_i|\widehat{I}_x|\psi_i\rangle,
\end{equation}
is determined by the Boltzmann factors $p_i = \exp(-E_i/k_BT)/\sum_i \exp(-E_i/k_BT)$ and
eigenvectors $\psi_i$ and eigenvalues $E_i$ of the nuclear Hamiltonian
\begin{equation}
    \begin{gathered}
    \widehat{H} = -g\mu_N\widehat{I_x}B_{\mathrm{hf}} + \frac{e^2qQ}{4I(2I-1)}(3\widehat{I}_z^2 - I(I+1)).
   \end{gathered}
\label{eq:Hamilton}
\end{equation}
For the quadrupolar parameters $eq$ and $Q$ and other details of this Hamiltonian, see main body and supplementary material of Ref.~\cite{Knapp2023}.

From thus determined hyperfine fields $b_{\mathrm{hf}}^{171}$ and $b_{\mathrm{hf}}^{173}$ and the natural abundances $n_{171} = 0.14$ and $n_{173} = 0.16$ of the magnetically active Yb isotopes
we compute the average hyperfine field across all Yb sites
$$\langle b_{\mathrm{hf}}\rangle = n_{171}b_{\mathrm{hf}}^{171}+n_{173}b_{\mathrm{hf}}^{173}.$$

\section{Effective Field in Spin 1/2 Model}

Here we consider the simple model with $N$ Yb sites, $N_n$ of which host nuclear spins 1/2,
and the remaining nucei are spinless~\cite[Eqs.~(S15)-(S23)]{Knapp2023}.
We restrict ourselves to AFM1 and PM phases, where
there is no significant staggered magnetization,
so $\mu_e$ coincides with the paramagnetic moment (in Ref.~\cite{Knapp2023} this is called $\mu_P$).
In the strongly-correlated electron system of YbRh$_2$Si$_2$
we assume $\mu_e$ to be the same on Yb sites with and without nuclear spins.
Then the Gibbs free energy is
\begin{align}
    G(T, B_{\mathrm{ext}}; \mu_e) = &-N_n k_B T \log\left(2\cosh\frac{g\mu_N A_{\text{hf}} \mu_e}{2k_BT}\right)\notag\\
    &+ N[\beta\mu_e^2 - B_{\mathrm{ext}}\mu_{e}].\label{eq:Gibbs:muP}
\end{align}
For each site with a non-zero nuclear spin the thermal average of the nuclear polarization is
\begin{equation}\label{eq:Ix}
\langle I_x \rangle = -\frac{1}{2} \tanh \frac{g\mu_NA_{\text{hf}}\mu_e}{2k_BT},
\end{equation}
and after averaging over the Yb sites with and without nuclear spins we get
the effective hyperfine field
\begin{equation}\label{eq:Bhf:spinhalf}
\langle b_{\text{hf}} \rangle = \frac{N_n}{N}\frac{g\mu_NA_{\text{hf}}}{2}\tanh \frac{g\mu_NA_{\text{hf}}\mu_e}{2k_BT}.
\end{equation}
The equilibrium $\mu_e(T, B)$ is determined by minimising the free energy given by Eq.~\eqref{eq:Gibbs:muP},
\begin{align}
    \frac{\partial G}{\partial \mu_e} = &-N_n\frac{g\mu_NA_{\text{hf}}}{2}\tanh \frac{g\mu_NA_{\text{hf}}\mu_e}{2k_BT}\notag\\
    & + 2N\beta\mu_e - N B_{\mathrm{ext}} = 0.\label{eq:muP1}
\end{align}
This condition can be rewritten as
\begin{equation}\label{eq:muP}
    \frac{\partial G}{\partial \mu_e} = N\big(2\beta\mu_e - B_{\mathrm{ext}} - \langle b_{\text{hf}} \rangle\big) = 0.
\end{equation}
The solution of Eq.~\eqref{eq:muP} is
\begin{equation}\mu_e = [B_{\mathrm{ext}} + \langle b_{\text{hf}} \rangle] / 2\beta = B^* / 2\beta,
\label{eq:mu:Beff}
\end{equation}
thus the paramagnetic moments $\mu_e$ are driven by the effective field $B^*$ in a temperature-independent way.

\section{Effective Magnetization Curve}

Down to 13\,mK we neglect the hyperfine field
$\langle b_{\mathrm{hf}}\rangle$, which is at most 1\,mT,
thus in this regime $B^* \approx B_{\text{ext}}$.
At 13--40\,mK the temperature dependence of the magnetic susceptibility $\chi_m(B_{\text{ext}})$
of the AFM1 phase~\cite{Steinke2017},
and therefore of the magnetization curve $\mu_e(B_{\text{ext}})$ is weak, consistent with temperature-independent $\mu_e(B^*)$
derived in Eq.~\eqref{eq:mu:Beff}.
We assume $\mu_e(B^*)$ to have this property,
and derive this dependence from the 13\,mK
susceptibility measurements~\cite{Steinke2017} as
\begin{align}
\mu_e(B^*, T) &\approx \mu_e(B_{\text{ext}}, 13\,\mathrm{mK})\notag\\
&= \int\limits_{B = 0}^{B_{\text{ext}}} \frac{1.05\chi_m(B, 13\,\mathrm{mK})}{\mu_0 N_A} \, dB 
\label{eq:chi:eff}
\end{align}
We apply the 1.05 scaling factor to the molar susceptibility $\chi_m(B, T)$ from Ref.~\cite{Steinke2017},
since this yields a good match with $\mu_e(B^*)$ derived
from heat capacity, see Fig.~2 of the main text and the next section.

\section{Hyperfine Schottky Model}

\begin{figure}
    \centering
    \includegraphics[width=2.96in]{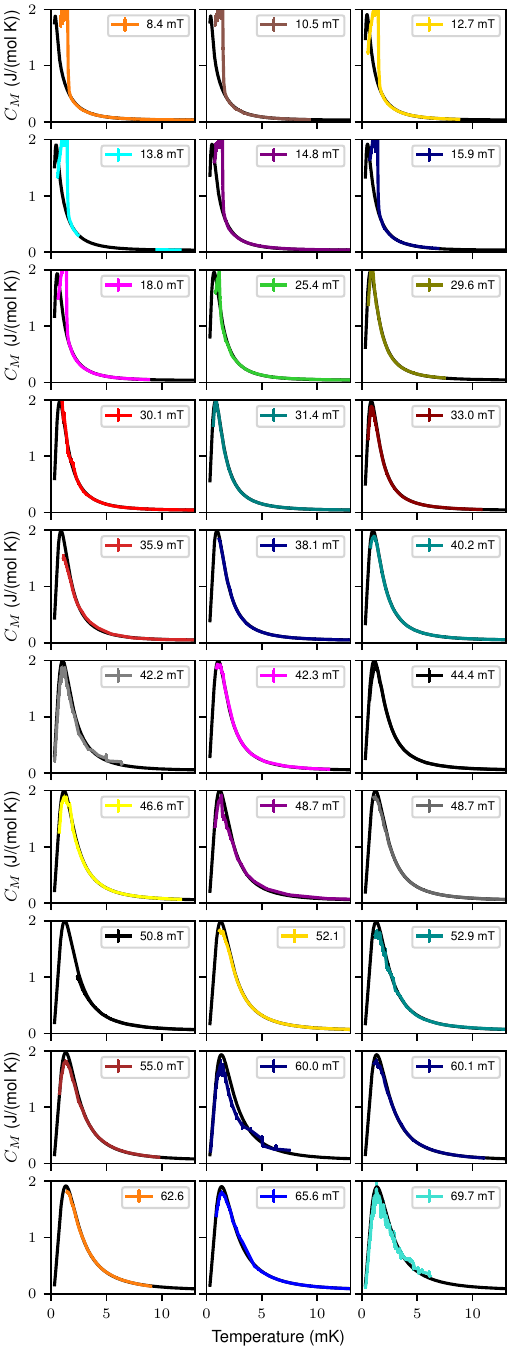}
    \caption{
    Fitting hyperfine Schottky model to data at various external fields.
    The fits are shown by the black lines.
    The fitting interval is 1.7--13\,mK.
    In fields up to 34\,mT, we observe a discrepancy between the data and the model at low temperature associated with
    the AFM1/AFM2 transition.
    At higher fields the $T < 1.7$\,mK data systematically undercut the model. This small discrepancy of unknown origin
    demonstrates the increasing complexity of YbRh$_2$Si$_2$ as $T\rightarrow 0$.}
    \label{fig:Schottky_panel}
\end{figure}

Fig.~2 in the main body shows a detailed study of the growth of the electronic moment with external magnetic field inside the AFM1 order and across the AFM1/PM boundary.
Our previous discussion of this dependence in Ref.~\cite{Knapp2023} (see Fig.~2 there), neglected the effect of the hyperfine field $b_{\mathrm{hf}}$.
Here we show how it can be taken into account
in the analysis of the heat capacity measurements.
The fitting at 1.7--13\,mK of the high-temperature ``tails''
of the Schottky-like heat capacity peaks is illustrated in Fig.~\ref{fig:Schottky_panel}.

The \emph{Hyperfine Schottky model} 
considers the exposure of the nuclear magnetic moments to the hyperfine field $B_{\mathrm{hf}}(T) = -A_{\mathrm{hf}}\mu_e(T)$, where the value of the electronic moment $\mu_e(T)$ in AFM1 increases as temperature is lowered from
the high temperature value $\mu_e^{\infty} \approx \mu_e(13\,\mathrm{mK})$.
This is a result of the temperature-dependent hyperfine field $\langle b_{\mathrm{hf}}\rangle$ acting additionally on the electronic moment.

At a fixed $B_{\text{ext}}$ we calculate the heat capacity
$C_M = T \partial S_M/\partial T$ from the entropy $S_M = -k_B\sum_{i=-I}^I p_i\log p_i$,
which is a function of the thermodynamic state 
fully determined by $B_{\mathrm{hf}}(T)$ and $T$.
In the fitting routine $\mu_e(T)$ is found at an arbitrary temperature $T$ by solving the equation $\mu_e(T) - \mu_e^\infty - \chi_{\mathrm{ion}}\langle b_{\mathrm{hf}}(T, \mu_e(T))\rangle = 0$
treating $\mu_e^\infty$ as the fitting parameter.

The value of $\chi_{\mathrm{ion}} = \partial \mu_e^{\infty} / \partial B_{\mathrm{ext}}$ is itself field dependent.
Here, we obtain it, similar to Eq.~\eqref{eq:chi:eff}, from the 13\,mK susceptibility data~\cite{Steinke2017} as $\chi_{\mathrm{ion}}(B_{\text{ext}}) = 1.05\chi_m(B_{\text{ext}}, 13\,\mathrm{mK}) \big / \mu_0 N_A$,
leading to a good agreement between the two methods
of determining $\mu_e^{\infty}(B_{\text{ext}})$, see Fig.~2.
One may also derive $\chi_{\mathrm{ion}}$ from the heat capacity data simultaneously with the determination of $\mu_e^{\infty}(B_{\text{ext}})$, but this procedure is computationally complex and may suffer from noise in the determined 
$\mu_e^{\infty}(B_{\text{ext}})$.


Our fits take into account  the heat capacity \mbox{$C = \gamma T$} of the heavy fermi liquid below 57\,mT and $C = a\log(T_0/T)T$ due to quantum-critical quasiparticles at higher fields.
The field dependence of the Sommerfeld coefficient $\gamma$, constant $a$, and characteristic temperature $T_0$
will be the subject of a future report.

\begin{figure}[t!]
    \centering
    \includegraphics{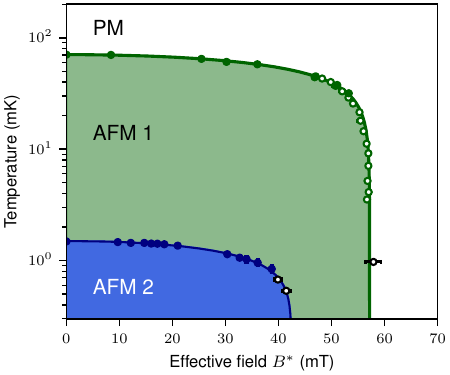}
    \caption{The phase diagram of YbRh\textsubscript{2}Si\textsubscript{2} as a function of effective field $B^*$ and temperature.
    In these coordinates both phase boundaries have conventional shape with no backturn.
    The significance of $B^*$ inside the AFM2 dome remains an open question.
    }
    \label{fig:Phase_diagram_B_effective}
\end{figure}

\begin{figure}[!b]
    \centering
    \includegraphics{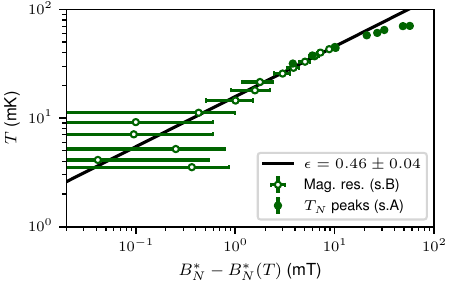}
    \caption{AFM1/PM transition points in the vicinity of the critical end point.
    The approach is with infinite slope and the exponent is $\varepsilon$ close to 1/2.
    }
    \label{fig:Phase_diagram_B_reduced}
\end{figure}

\section{Magneto-caloric effect in Y\lowercase{b}R\lowercase{h}$_{\mathbf{2}}$S\lowercase{i}$_{\mathrm{2}}$}

In an ideal isentropic magneto-caloric sweep on a simple non-interacting spin system, $B/T$ remains constant~\cite{Andres1982}.
In our system, the dominant magnetic field felt by the active Yb nuclei is the hyperfine field $B_{\mathrm{hf}} = -A_{\mathrm{hf}}\mu_e$.
Disregarding the quadrupolar interactions, 
$B_{\mathrm{hf}} / T$ and therefore 
 $\mu_e / T$ are constant along the sweep, and the shape of the $T(B_{\mathrm{ext}})$ curve directly reflects $\mu_e(B_{\mathrm{ext}})$.
The subtlety is that $\mu_e$ needs to be inferred along the $T(B_{\mathrm{ext}})$ trajectory.
The conservation of the nuclear entropy during an isentropic sweep implies that the nuclear polarization and hence the hyperfine field $\langle b_{\mathrm{hf}}\rangle$ remains constant. Thus $T(B_{\mathrm{ext}})$ is a straight line over any field range where the susceptibility is field-independent: here $T \propto \mu_e = \chi_{\mathrm{ion}} B^* = \chi_{\mathrm{ion}}\big(B_{\mathrm{ext}} + \langle b_{\mathrm{hf}}\rangle\big)$. 

The quadrupolar interactions on $^{173}$Yb can be taken into account numerically.
Despite small perturbations to the behavior described above, the key qualitative results hold:
$T(B_{\mathrm{ext}})$ is approximately linear when $\chi$ is field-independent
and a change in $\chi$ leads to a kink in $T(B_{\mathrm{ext}})$.


\begin{figure}[!b]
    \centering
    \includegraphics{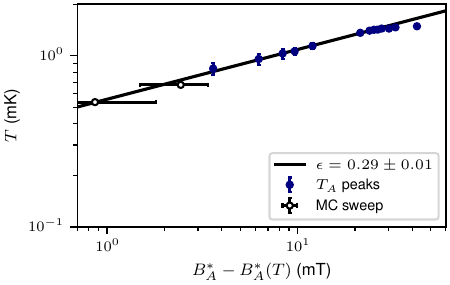}
    \caption{AFM1/AFM2 transition points in the vicinity of the critical end point.
    The approach is with infinite slope and the exponent is $\varepsilon = 0.29 \pm 0.01$.
    }
    \label{fig:Phase_diagram_B_reduced_AFM2}
\end{figure}

Examining Fig.~4b, we can see that inside the AFM2 phase the temperature increases during field sweeps in both directions.
This 
is a consequence of the parasitic heat deposition during the sweeps. Correspondingly, the slope of the magnetization sweep is steeper than that of the isentropic sweep.
In future measurements this effect can be corrected for if demagnetization and magnetization sweeps are taken under similar conditions, 
back-to-back, 
and the temperature points in matching fields on the way down and up are averaged together.


\begin{figure*}[t!]
    \parbox[t]{3.4in}{\centering
    \includegraphics{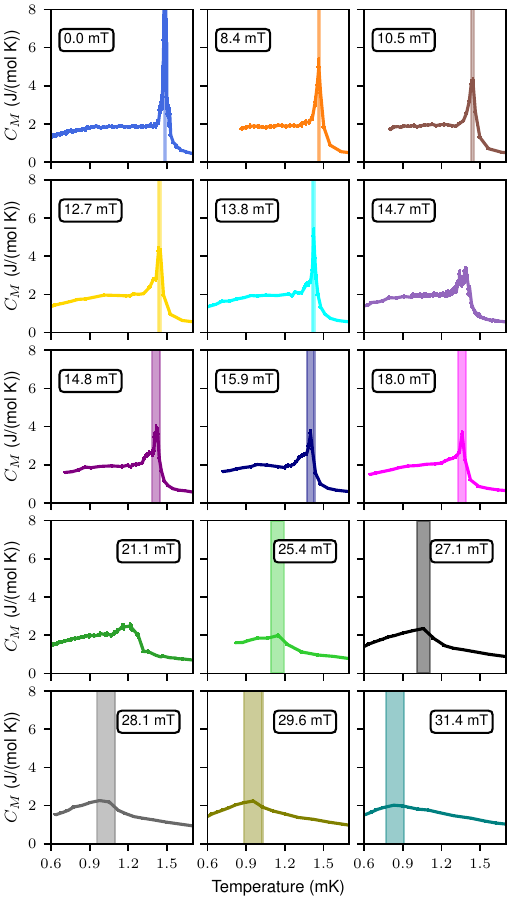}
    \caption{Evolution of the $T_A$ transition peak with external magnetic field.
    Estimated $T$ position and uncertainty marked by the band.
    The ``final field cooled'' measurements above 11\,mT (14.7\,mT and 21.1\,mT) are excluded from the phase diagram shown in Fig.~1.
    }
    \label{fig:TA_peaks}}\hfill%
    \parbox[t]{3.4in}{\centering
    \includegraphics{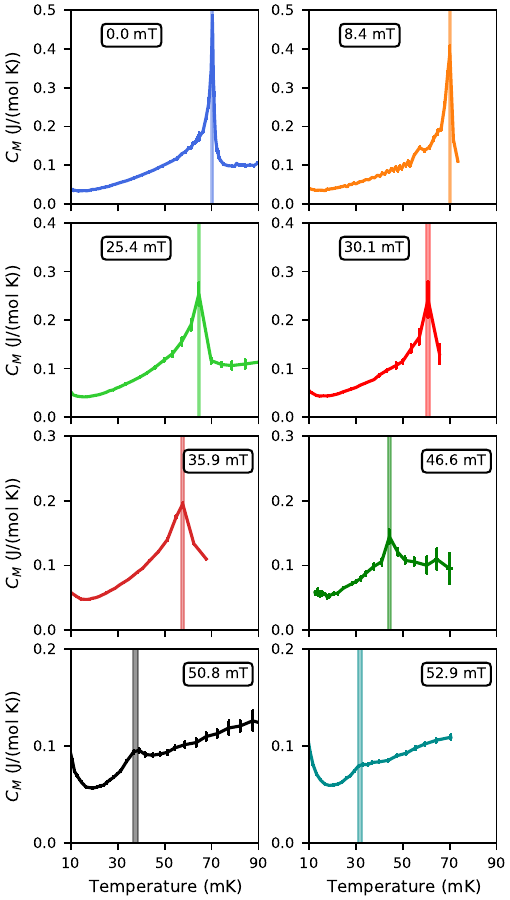}
    \caption{Evolution of the $T_N$ transition peak with external magnetic field.
    Estimated $T$ position and uncertainty marked by the band.
    }
    \label{fig:TN_peaks}}
\end{figure*}

\section{AFM1/PM Phase Boundary}

We consider an ansatz 
\begin{equation}
    B_N^*(T) = \sqrt{a_3T^3 + a_2T^2 + a_1T + a_0},
\label{eq:Cube}
\end{equation}
for the effective critical field of AFM1.
Taking $\mu_e(B^*)$ from Eq.~\eqref{eq:chi:eff} we derive the external critical field
$$
     B_N(T) = B_N^*(T) - \big\langle b_{\mathrm{hf}}\big(T,\mu_e(B_N^*(T))\big)\big\rangle
$$
and fit it to the experimental signatures of the AFM1/PM phase boundary, shown in Fig.~1. This procedure yields
\medskip

\centerline{\begin{tabular}{c|c}
    $a_0$ & $3.239\cdot10^{3}$\,mT\textsuperscript{2}, \\
    $a_1$ & $4.281\cdot10^{-1}$\,mT\textsuperscript{2}/mK, \\
    $a_2$ & $-3.110\cdot10^{-1}$\,mT\textsuperscript{2}/mK\textsuperscript{2}, \\
    $a_3$ & $-4.946\cdot10^{-3}$\,mT\textsuperscript{2}/mK\textsuperscript{3}. \\
\end{tabular}}
\medskip

\noindent We note that this model involves no free parameters to describe the hyperfine field responsible for the backturn below 10\,mK.

We also evaluate the effective field for the experimental signatures of this phase boundary, as shown in Fig.~\ref{fig:Phase_diagram_B_effective}, and observe
a conventional dome with no backturn. 
This supports the hypothesis
that the antiferromagnetism in YbRh$_2$Si$_2$ is driven by the  effective field $B^*$. 

Fig.~\ref{fig:Phase_diagram_B_reduced} studies the vicinity of the critical end point 
$B_N^\ast = 57.05\pm0.3$\,mT.
A power law fit $T = (B_N^\ast - B_N^\ast(T))^{\varepsilon}$ yields $\varepsilon = 0.46\pm0.04$.
Only the magneto-resistance signatures are analyzed, in order to avoid systematic errors from combining the data from different experimental setups.
The $\pm 0.5$\,mT uncertainty on these $B_N^*(T)$ measurements renders the data below 14\,mK insignificant to this analysis, therefore effectively we determine $\varepsilon$
over the 14-45\,mK temperature range, equivalent to
the reduced temperature range $T/T_N(B=0) = 0.2$-0.6. More accurate determination of $B_N^*(T)$ is desirable in order to better access the behavior in the $T \to 0$ limit. Furthermore, 
errors associated with the $\langle b_{\mathrm{hf}}\rangle$ subtraction,
can be eliminated in an experiment with no Yb nuclear spins, i.e.\ on $^{174}$YbRh$_2$Si$_2$, where we expect $B_N^*(T)$ to be measured directly. 

\begin{figure}[!b]
    \centering
    \includegraphics[width=0.7\linewidth]{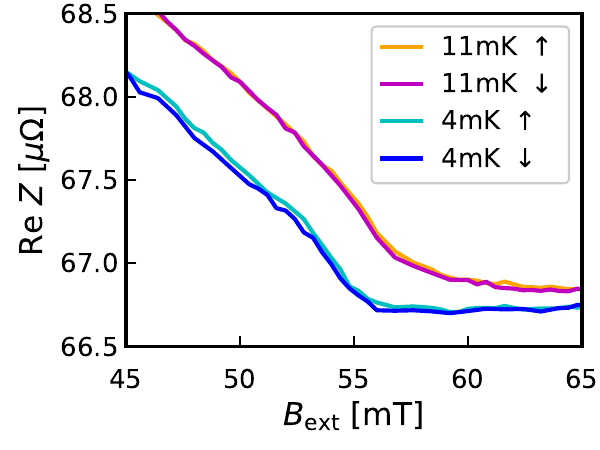}
    \caption{Magnetoresistance as a function of external field swept up and down (direction indicated by the arrows).}
    \label{fig:MR_up_and_down}
\end{figure}


Our value of $\varepsilon$ is larger than previously reported $\varepsilon = 0.36$~\cite{Friedemann2009} inferred from the 20-40\,mK range. Over this limited temperature range our data are also consistent with $\varepsilon = 0.36$, emphasizing the importance of measurements at lower temperatures.

\section{AFM1/AFM2 Phase Boundary}

The phase boundary of AFM2 is described well by a similar ansatz,
\begin{equation}
B_A(T) = \sqrt{ A_3T^3 + A_2T^2 + A_1T + A_0 },
\end{equation}
and the best fit parameters are
\medskip

\centerline{\begin{tabular}{c|c}
    $A_0$ & $1.096\cdot10^{3}$\,mT\textsuperscript{2}, \\
    $A_1$ & $-1.112\cdot10^{2}$\,mT\textsuperscript{2}/mK, \\
    $A_2$ & $3.528\cdot10^{2}$\,mT\textsuperscript{2}/mK\textsuperscript{2}, \\
    $A_3$ & $-5.120\cdot10^{2}$\,mT\textsuperscript{2}/mK\textsuperscript{3}. \\
\end{tabular}}
\medskip

\noindent Similar to $B_N^*(T)$ the AFM1/AFM2 phase boundary
can be considered in the temperature-effective field plane,
see Fig.~\ref{fig:Phase_diagram_B_effective}.
For this phase boundary a conventional dome is observed, consistent with the AFM2 order being suppressed by the effective field $B^*$.
Nevertheless, the dome has similar shape in $B_A(T)$ and $B_A^*(T)$ coordinates, as $\langle b_{\mathrm{hf}}\rangle$ is saturated along a significant part of this phase boundary.

The approach of the critical end point of AFM2, $B_A^\ast$, again in terms of effective field, is studied in Fig.~\ref{fig:Phase_diagram_B_reduced}.
A power law fit $T = (B_A^\ast - B_A^\ast(T))^{\varepsilon}$ yields $\varepsilon = 0.29 \pm 0.01$, different to the AFM1 dome.
Here the data are limited to $T > 0.54\,\mathrm{mK} = 0.36T_A(B=0)$, and measurements at lower temperatures may be necessary to determine the $T \to 0$ behavior.

In low fields both phase transition temperatures become nearly field independent, $\partial T_{A}/\partial B_{\mathrm{ext}}=0$ and $\partial T_{N}/\partial B_{\mathrm{ext}}=0$ at $B_{\mathrm{ext}} = 0$.
This is in contrast to the finite $\partial T_{A}/\partial B_{\mathrm{ext}}$ reported from magnetic susceptibility measurements~\cite{Schuberth2016};
we note that those studies would have been affected by the diamagnetic screening associated with superconductivity.

\section{AFM1/AFM2 Transition}

The complete sets of heat capacity peaks at AFM1/AFM2 and PM/AFM1 phase transitions are shown in Figs.~\ref{fig:TA_peaks} and \ref{fig:TN_peaks} respectively.
Although both transitions appear to be of second order at zero magnetic field~\cite{Krellner2009,Knapp2023,Knapp2024}, there is evidence for weak dependence of $T_A$ signatures on the trajectory
in $(T, B_{\mathrm{ext}})$ plane prior to crossing the phase boundary.
In particular, a double peak is observed at 14.7\,mT (Fig.~\ref{fig:TA_peaks}) when the sample is ``final field cooled'', i.e.\ cooled and subsequently measured at the same field (this is the natural approach in the experiment with aluminium heat switch above its switching field of 10\,mT~\cite{Knapp2024}).
In contrast, when the sample is cooled in 80\,mT with the field lowered after reaching the base temperature (required with the lead heat switch, which has a higher switching field~\cite{Knapp2024}), the double peak effectively disappears (Fig.~\ref{fig:TA_peaks}, 14.8\,mT) and the AFM2 order survives to a slightly higher temperature on warming.
This history effect may be linked to how the AFM1 and AFM2 magnetic domains respond to magnetic field.
No history effects were observed below 11\,mT; 
at higher fields  ``final field cooled'' measurements are excluded from the  data analysis. 

\section{Hysteresis in magnetoresistance?}

To support the assertion of the AFM1/PM transition being second order, we checked for hysteresis when crossing the boundary in opposite directions of the field sweep.
The result is shown in Fig.~\ref{fig:MR_up_and_down} and indicates no hysteresis.


\begin{thebibliography}{55}%
\makeatletter
\providecommand \@ifxundefined [1]{%
 \@ifx{#1\undefined}
}%
\providecommand \@ifnum [1]{%
 \ifnum #1\expandafter \@firstoftwo
 \else \expandafter \@secondoftwo
 \fi
}%
\providecommand \@ifx [1]{%
 \ifx #1\expandafter \@firstoftwo
 \else \expandafter \@secondoftwo
 \fi
}%
\providecommand \natexlab [1]{#1}%
\providecommand \enquote  [1]{``#1''}%
\providecommand \bibnamefont  [1]{#1}%
\providecommand \bibfnamefont [1]{#1}%
\providecommand \citenamefont [1]{#1}%
\providecommand \href@noop [0]{\@secondoftwo}%
\providecommand \href [0]{\begingroup \@sanitize@url \@href}%
\providecommand \@href[1]{\@@startlink{#1}\@@href}%
\providecommand \@@href[1]{\endgroup#1\@@endlink}%
\providecommand \@sanitize@url [0]{\catcode `\\12\catcode `\$12\catcode `\&12\catcode `\#12\catcode `\^12\catcode `\_12\catcode `\%12\relax}%
\providecommand \@@startlink[1]{}%
\providecommand \@@endlink[0]{}%
\providecommand \url  [0]{\begingroup\@sanitize@url \@url }%
\providecommand \@url [1]{\endgroup\@href {#1}{\urlprefix }}%
\providecommand \urlprefix  [0]{URL }%
\providecommand \Eprint [0]{\href }%
\providecommand \doibase [0]{https://doi.org/}%
\providecommand \selectlanguage [0]{\@gobble}%
\providecommand \bibinfo  [0]{\@secondoftwo}%
\providecommand \bibfield  [0]{\@secondoftwo}%
\providecommand \translation [1]{[#1]}%
\providecommand \BibitemOpen [0]{}%
\providecommand \bibitemStop [0]{}%
\providecommand \bibitemNoStop [0]{.\EOS\space}%
\providecommand \EOS [0]{\spacefactor3000\relax}%
\providecommand \BibitemShut  [1]{\csname bibitem#1\endcsname}%
\let\auto@bib@innerbib\@empty
\bibitem [{\citenamefont {Gegenwart}\ \emph {et~al.}(2002)\citenamefont {Gegenwart}, \citenamefont {Custers}, \citenamefont {Geibel}, \citenamefont {Neumaier}, \citenamefont {Tayama}, \citenamefont {Tenya}, \citenamefont {Trovarelli},\ and\ \citenamefont {Steglich}}]{Gegenwart2002}%
  \BibitemOpen
  \bibfield  {author} {\bibinfo {author} {\bibfnamefont {P.}~\bibnamefont {Gegenwart}}, \bibinfo {author} {\bibfnamefont {J.}~\bibnamefont {Custers}}, \bibinfo {author} {\bibfnamefont {C.}~\bibnamefont {Geibel}}, \bibinfo {author} {\bibfnamefont {K.}~\bibnamefont {Neumaier}}, \bibinfo {author} {\bibfnamefont {T.}~\bibnamefont {Tayama}}, \bibinfo {author} {\bibfnamefont {K.}~\bibnamefont {Tenya}}, \bibinfo {author} {\bibfnamefont {O.}~\bibnamefont {Trovarelli}},\ and\ \bibinfo {author} {\bibfnamefont {F.}~\bibnamefont {Steglich}},\ }\bibfield  {title} {\bibinfo {title} {Magnetic-field induced quantum critical point in {YbRh$_2$Si$_2$}},\ }\href {https://doi.org/10.1103/physrevlett.89.056402} {\bibfield  {journal} {\bibinfo  {journal} {Physical Review Letters}\ }\textbf {\bibinfo {volume} {89}},\ \bibinfo {pages} {056402} (\bibinfo {year} {2002})}\BibitemShut {NoStop}%
\bibitem [{\citenamefont {Custers}\ \emph {et~al.}(2003)\citenamefont {Custers}, \citenamefont {Gegenwart}, \citenamefont {Neumaier}, \citenamefont {Wilhelm}, \citenamefont {Oeschler}, \citenamefont {Ishida}, \citenamefont {Kitaoka}, \citenamefont {Geibel},\ and\ \citenamefont {Steglich}}]{Custers2003}%
  \BibitemOpen
  \bibfield  {author} {\bibinfo {author} {\bibfnamefont {J.}~\bibnamefont {Custers}}, \bibinfo {author} {\bibfnamefont {P.}~\bibnamefont {Gegenwart}}, \bibinfo {author} {\bibfnamefont {K.}~\bibnamefont {Neumaier}}, \bibinfo {author} {\bibfnamefont {H.}~\bibnamefont {Wilhelm}}, \bibinfo {author} {\bibfnamefont {N.}~\bibnamefont {Oeschler}}, \bibinfo {author} {\bibfnamefont {K.}~\bibnamefont {Ishida}}, \bibinfo {author} {\bibfnamefont {Y.}~\bibnamefont {Kitaoka}}, \bibinfo {author} {\bibfnamefont {C.}~\bibnamefont {Geibel}},\ and\ \bibinfo {author} {\bibfnamefont {F.}~\bibnamefont {Steglich}},\ }\bibfield  {title} {\bibinfo {title} {Quantum criticality in {YbRh}$_2${Si}$_2$},\ }\href {https://doi.org/10.1088/0953-8984/15/28/324} {\bibfield  {journal} {\bibinfo  {journal} {Journal of Physics: Condensed Matter}\ }\textbf {\bibinfo {volume} {15}},\ \bibinfo {pages} {S2047} (\bibinfo {year} {2003})}\BibitemShut {NoStop}%
\bibitem [{\citenamefont {Pfau}\ \emph {et~al.}(2012)\citenamefont {Pfau}, \citenamefont {Hartmann}, \citenamefont {Stockert}, \citenamefont {Sun}, \citenamefont {Lausberg}, \citenamefont {Brando}, \citenamefont {Friedemann}, \citenamefont {Krellner}, \citenamefont {Geibel}, \citenamefont {Wirth}, \citenamefont {Kirchner}, \citenamefont {Abrahams}, \citenamefont {Si},\ and\ \citenamefont {Steglich}}]{Pfau2012}%
  \BibitemOpen
  \bibfield  {author} {\bibinfo {author} {\bibfnamefont {H.}~\bibnamefont {Pfau}}, \bibinfo {author} {\bibfnamefont {S.}~\bibnamefont {Hartmann}}, \bibinfo {author} {\bibfnamefont {U.}~\bibnamefont {Stockert}}, \bibinfo {author} {\bibfnamefont {P.}~\bibnamefont {Sun}}, \bibinfo {author} {\bibfnamefont {S.}~\bibnamefont {Lausberg}}, \bibinfo {author} {\bibfnamefont {M.}~\bibnamefont {Brando}}, \bibinfo {author} {\bibfnamefont {S.}~\bibnamefont {Friedemann}}, \bibinfo {author} {\bibfnamefont {C.}~\bibnamefont {Krellner}}, \bibinfo {author} {\bibfnamefont {C.}~\bibnamefont {Geibel}}, \bibinfo {author} {\bibfnamefont {S.}~\bibnamefont {Wirth}}, \bibinfo {author} {\bibfnamefont {S.}~\bibnamefont {Kirchner}}, \bibinfo {author} {\bibfnamefont {E.}~\bibnamefont {Abrahams}}, \bibinfo {author} {\bibfnamefont {Q.}~\bibnamefont {Si}},\ and\ \bibinfo {author} {\bibfnamefont {F.}~\bibnamefont {Steglich}},\ }\bibfield  {title} {\bibinfo {title} {Thermal and electrical transport across a magnetic quantum critical point},\
  }\href {https://doi.org/10.1038/nature11072} {\bibfield  {journal} {\bibinfo  {journal} {Nature}\ }\textbf {\bibinfo {volume} {484}},\ \bibinfo {pages} {493} (\bibinfo {year} {2012})}\BibitemShut {NoStop}%
\bibitem [{\citenamefont {Gegenwart}\ \emph {et~al.}(2007)\citenamefont {Gegenwart}, \citenamefont {Westerkamp}, \citenamefont {Krellner}, \citenamefont {Tokiwa}, \citenamefont {Paschen}, \citenamefont {Geibel}, \citenamefont {Steglich}, \citenamefont {Abrahams},\ and\ \citenamefont {Si}}]{Gegenwart2007}%
  \BibitemOpen
  \bibfield  {author} {\bibinfo {author} {\bibfnamefont {P.}~\bibnamefont {Gegenwart}}, \bibinfo {author} {\bibfnamefont {T.}~\bibnamefont {Westerkamp}}, \bibinfo {author} {\bibfnamefont {C.}~\bibnamefont {Krellner}}, \bibinfo {author} {\bibfnamefont {Y.}~\bibnamefont {Tokiwa}}, \bibinfo {author} {\bibfnamefont {S.}~\bibnamefont {Paschen}}, \bibinfo {author} {\bibfnamefont {C.}~\bibnamefont {Geibel}}, \bibinfo {author} {\bibfnamefont {F.}~\bibnamefont {Steglich}}, \bibinfo {author} {\bibfnamefont {E.}~\bibnamefont {Abrahams}},\ and\ \bibinfo {author} {\bibfnamefont {Q.}~\bibnamefont {Si}},\ }\bibfield  {title} {\bibinfo {title} {Multiple energy scales at a quantum critical point},\ }\href {https://doi.org/10.1126/science.1136020} {\bibfield  {journal} {\bibinfo  {journal} {Science}\ }\textbf {\bibinfo {volume} {315}},\ \bibinfo {pages} {969} (\bibinfo {year} {2007})}\BibitemShut {NoStop}%
\bibitem [{\citenamefont {Gegenwart}\ \emph {et~al.}(2008)\citenamefont {Gegenwart}, \citenamefont {Si},\ and\ \citenamefont {Steglich}}]{Gegenwart2008}%
  \BibitemOpen
  \bibfield  {author} {\bibinfo {author} {\bibfnamefont {P.}~\bibnamefont {Gegenwart}}, \bibinfo {author} {\bibfnamefont {Q.}~\bibnamefont {Si}},\ and\ \bibinfo {author} {\bibfnamefont {F.}~\bibnamefont {Steglich}},\ }\bibfield  {title} {\bibinfo {title} {Quantum criticality in heavy{\textendash}fermion metals},\ }\href {https://doi.org/10.1038/nphys892} {\bibfield  {journal} {\bibinfo  {journal} {Nature Physics}\ }\textbf {\bibinfo {volume} {4}},\ \bibinfo {pages} {186} (\bibinfo {year} {2008})}\BibitemShut {NoStop}%
\bibitem [{\citenamefont {Friedemann}\ \emph {et~al.}(2009)\citenamefont {Friedemann}, \citenamefont {Westerkamp}, \citenamefont {Brando}, \citenamefont {Oeschler}, \citenamefont {Wirth}, \citenamefont {Gegenwart}, \citenamefont {Krellner}, \citenamefont {Geibel},\ and\ \citenamefont {Steglich}}]{Friedemann2009}%
  \BibitemOpen
  \bibfield  {author} {\bibinfo {author} {\bibfnamefont {S.}~\bibnamefont {Friedemann}}, \bibinfo {author} {\bibfnamefont {T.}~\bibnamefont {Westerkamp}}, \bibinfo {author} {\bibfnamefont {M.}~\bibnamefont {Brando}}, \bibinfo {author} {\bibfnamefont {N.}~\bibnamefont {Oeschler}}, \bibinfo {author} {\bibfnamefont {S.}~\bibnamefont {Wirth}}, \bibinfo {author} {\bibfnamefont {P.}~\bibnamefont {Gegenwart}}, \bibinfo {author} {\bibfnamefont {C.}~\bibnamefont {Krellner}}, \bibinfo {author} {\bibfnamefont {C.}~\bibnamefont {Geibel}},\ and\ \bibinfo {author} {\bibfnamefont {F.}~\bibnamefont {Steglich}},\ }\bibfield  {title} {\bibinfo {title} {Detaching the antiferromagnetic quantum critical point from the {Fermi}-surface reconstruction in {YbRh\textsubscript{2}Si\textsubscript{2}}},\ }\href {https://doi.org/10.1038/nphys1299} {\bibfield  {journal} {\bibinfo  {journal} {Nature Physics}\ }\textbf {\bibinfo {volume} {5}},\ \bibinfo {pages} {465} (\bibinfo {year} {2009})}\BibitemShut {NoStop}%
\bibitem [{\citenamefont {Knebel}\ \emph {et~al.}(2006)\citenamefont {Knebel}, \citenamefont {Boursier}, \citenamefont {Hassinger}, \citenamefont {Lapertot}, \citenamefont {Niklowitz}, \citenamefont {Pourret}, \citenamefont {Salce}, \citenamefont {Sanchez}, \citenamefont {Sheikin}, \citenamefont {Bonville}, \citenamefont {Harima},\ and\ \citenamefont {Flouquet}}]{Knebel2006}%
  \BibitemOpen
  \bibfield  {author} {\bibinfo {author} {\bibfnamefont {G.}~\bibnamefont {Knebel}}, \bibinfo {author} {\bibfnamefont {R.}~\bibnamefont {Boursier}}, \bibinfo {author} {\bibfnamefont {E.}~\bibnamefont {Hassinger}}, \bibinfo {author} {\bibfnamefont {G.}~\bibnamefont {Lapertot}}, \bibinfo {author} {\bibfnamefont {P.~G.}\ \bibnamefont {Niklowitz}}, \bibinfo {author} {\bibfnamefont {A.}~\bibnamefont {Pourret}}, \bibinfo {author} {\bibfnamefont {B.}~\bibnamefont {Salce}}, \bibinfo {author} {\bibfnamefont {J.~P.}\ \bibnamefont {Sanchez}}, \bibinfo {author} {\bibfnamefont {I.}~\bibnamefont {Sheikin}}, \bibinfo {author} {\bibfnamefont {P.}~\bibnamefont {Bonville}}, \bibinfo {author} {\bibfnamefont {H.}~\bibnamefont {Harima}},\ and\ \bibinfo {author} {\bibfnamefont {J.}~\bibnamefont {Flouquet}},\ }\bibfield  {title} {\bibinfo {title} {Localization of 4f state in {YbRh}$_2${Si}$_2$ under magnetic field and high pressure: Comparison with {CeRh}$_2${Si}$_2$},\ }\href {https://doi.org/10.1143/jpsj.75.114709} {\bibfield
  {journal} {\bibinfo  {journal} {Journal of the Physical Society of Japan}\ }\textbf {\bibinfo {volume} {75}},\ \bibinfo {pages} {114709} (\bibinfo {year} {2006})}\BibitemShut {NoStop}%
\bibitem [{\citenamefont {Weng}\ \emph {et~al.}(2016)\citenamefont {Weng}, \citenamefont {Smidman}, \citenamefont {Jiao}, \citenamefont {Lu},\ and\ \citenamefont {Yuan}}]{Weng2016}%
  \BibitemOpen
  \bibfield  {author} {\bibinfo {author} {\bibfnamefont {Z.~F.}\ \bibnamefont {Weng}}, \bibinfo {author} {\bibfnamefont {M.}~\bibnamefont {Smidman}}, \bibinfo {author} {\bibfnamefont {L.}~\bibnamefont {Jiao}}, \bibinfo {author} {\bibfnamefont {X.}~\bibnamefont {Lu}},\ and\ \bibinfo {author} {\bibfnamefont {H.~Q.}\ \bibnamefont {Yuan}},\ }\bibfield  {title} {\bibinfo {title} {Multiple quantum phase transitions and superconductivity in ce-based heavy fermions},\ }\href {https://doi.org/10.1088/0034-4885/79/9/094503} {\bibfield  {journal} {\bibinfo  {journal} {Reports on Progress in Physics}\ }\textbf {\bibinfo {volume} {79}},\ \bibinfo {pages} {094503} (\bibinfo {year} {2016})}\BibitemShut {NoStop}%
\bibitem [{\citenamefont {Nakatsuji}\ \emph {et~al.}(2008)\citenamefont {Nakatsuji}, \citenamefont {Kuga}, \citenamefont {Machida}, \citenamefont {Tayama}, \citenamefont {Sakakibara}, \citenamefont {Karaki}, \citenamefont {Ishimoto}, \citenamefont {Yonezawa}, \citenamefont {Maeno}, \citenamefont {Pearson}, \citenamefont {Lonzarich}, \citenamefont {Balicas}, \citenamefont {Lee},\ and\ \citenamefont {Fisk}}]{Nakatsuji2008}%
  \BibitemOpen
  \bibfield  {author} {\bibinfo {author} {\bibfnamefont {S.}~\bibnamefont {Nakatsuji}}, \bibinfo {author} {\bibfnamefont {K.}~\bibnamefont {Kuga}}, \bibinfo {author} {\bibfnamefont {Y.}~\bibnamefont {Machida}}, \bibinfo {author} {\bibfnamefont {T.}~\bibnamefont {Tayama}}, \bibinfo {author} {\bibfnamefont {T.}~\bibnamefont {Sakakibara}}, \bibinfo {author} {\bibfnamefont {Y.}~\bibnamefont {Karaki}}, \bibinfo {author} {\bibfnamefont {H.}~\bibnamefont {Ishimoto}}, \bibinfo {author} {\bibfnamefont {S.}~\bibnamefont {Yonezawa}}, \bibinfo {author} {\bibfnamefont {Y.}~\bibnamefont {Maeno}}, \bibinfo {author} {\bibfnamefont {E.}~\bibnamefont {Pearson}}, \bibinfo {author} {\bibfnamefont {G.~G.}\ \bibnamefont {Lonzarich}}, \bibinfo {author} {\bibfnamefont {L.}~\bibnamefont {Balicas}}, \bibinfo {author} {\bibfnamefont {H.}~\bibnamefont {Lee}},\ and\ \bibinfo {author} {\bibfnamefont {Z.}~\bibnamefont {Fisk}},\ }\bibfield  {title} {\bibinfo {title} {Superconductivity and quantum criticality in the heavy-fermion system
  $\beta${\textendash}{YbAlB}$_4$},\ }\href {https://doi.org/10.1038/nphys1002} {\bibfield  {journal} {\bibinfo  {journal} {Nature Physics}\ }\textbf {\bibinfo {volume} {4}},\ \bibinfo {pages} {603} (\bibinfo {year} {2008})}\BibitemShut {NoStop}%
\bibitem [{\citenamefont {Schuberth}\ \emph {et~al.}(2016)\citenamefont {Schuberth}, \citenamefont {Tippmann}, \citenamefont {Steinke}, \citenamefont {Lausberg}, \citenamefont {Steppke}, \citenamefont {Brando}, \citenamefont {Krellner}, \citenamefont {Geibel}, \citenamefont {Yu}, \citenamefont {Si},\ and\ \citenamefont {Steglich}}]{Schuberth2016}%
  \BibitemOpen
  \bibfield  {author} {\bibinfo {author} {\bibfnamefont {E.}~\bibnamefont {Schuberth}}, \bibinfo {author} {\bibfnamefont {M.}~\bibnamefont {Tippmann}}, \bibinfo {author} {\bibfnamefont {L.}~\bibnamefont {Steinke}}, \bibinfo {author} {\bibfnamefont {S.}~\bibnamefont {Lausberg}}, \bibinfo {author} {\bibfnamefont {A.}~\bibnamefont {Steppke}}, \bibinfo {author} {\bibfnamefont {M.}~\bibnamefont {Brando}}, \bibinfo {author} {\bibfnamefont {C.}~\bibnamefont {Krellner}}, \bibinfo {author} {\bibfnamefont {C.}~\bibnamefont {Geibel}}, \bibinfo {author} {\bibfnamefont {R.}~\bibnamefont {Yu}}, \bibinfo {author} {\bibfnamefont {Q.}~\bibnamefont {Si}},\ and\ \bibinfo {author} {\bibfnamefont {F.}~\bibnamefont {Steglich}},\ }\bibfield  {title} {\bibinfo {title} {Emergence of superconductivity in the canonical heavy{\textendash}electron metal {YbRh}$_2${Si}$_2$},\ }\href {https://doi.org/10.1126/science.aaa9733} {\bibfield  {journal} {\bibinfo  {journal} {Science}\ }\textbf {\bibinfo {volume} {351}},\ \bibinfo {pages} {485}
  (\bibinfo {year} {2016})}\BibitemShut {NoStop}%
\bibitem [{\citenamefont {Nguyen}\ \emph {et~al.}(2021)\citenamefont {Nguyen}, \citenamefont {Sidorenko}, \citenamefont {Taupin}, \citenamefont {Knebel}, \citenamefont {Lapertot}, \citenamefont {Schuberth},\ and\ \citenamefont {Paschen}}]{Nguyen2021}%
  \BibitemOpen
  \bibfield  {author} {\bibinfo {author} {\bibfnamefont {D.}~\bibnamefont {Nguyen}}, \bibinfo {author} {\bibfnamefont {A.}~\bibnamefont {Sidorenko}}, \bibinfo {author} {\bibfnamefont {M.}~\bibnamefont {Taupin}}, \bibinfo {author} {\bibfnamefont {G.}~\bibnamefont {Knebel}}, \bibinfo {author} {\bibfnamefont {G.}~\bibnamefont {Lapertot}}, \bibinfo {author} {\bibfnamefont {E.}~\bibnamefont {Schuberth}},\ and\ \bibinfo {author} {\bibfnamefont {S.}~\bibnamefont {Paschen}},\ }\bibfield  {title} {\bibinfo {title} {Superconductivity in an extreme strange metal},\ }\bibfield  {journal} {\bibinfo  {journal} {Nature Communications}\ }\textbf {\bibinfo {volume} {12}},\ \href {https://doi.org/10.1038/s41467-021-24670-z} {10.1038/s41467-021-24670-z} (\bibinfo {year} {2021})\BibitemShut {NoStop}%
\bibitem [{\citenamefont {Levitin}\ \emph {et~al.}(2025)\citenamefont {Levitin}, \citenamefont {Knapp}, \citenamefont {Knappová}, \citenamefont {Lucas}, \citenamefont {Nyéki}, \citenamefont {Heikkinen}, \citenamefont {Antonov}, \citenamefont {Casey}, \citenamefont {Ho}, \citenamefont {Coleman}, \citenamefont {Geibel}, \citenamefont {Steppke}, \citenamefont {Kliemt}, \citenamefont {Krellner}, \citenamefont {Brando},\ and\ \citenamefont {Saunders}}]{Levitin2025}%
  \BibitemOpen
  \bibfield  {author} {\bibinfo {author} {\bibfnamefont {L.~V.}\ \bibnamefont {Levitin}}, \bibinfo {author} {\bibfnamefont {J.}~\bibnamefont {Knapp}}, \bibinfo {author} {\bibfnamefont {P.}~\bibnamefont {Knappová}}, \bibinfo {author} {\bibfnamefont {M.}~\bibnamefont {Lucas}}, \bibinfo {author} {\bibfnamefont {J.}~\bibnamefont {Nyéki}}, \bibinfo {author} {\bibfnamefont {P.}~\bibnamefont {Heikkinen}}, \bibinfo {author} {\bibfnamefont {V.}~\bibnamefont {Antonov}}, \bibinfo {author} {\bibfnamefont {A.}~\bibnamefont {Casey}}, \bibinfo {author} {\bibfnamefont {A.~F.}\ \bibnamefont {Ho}}, \bibinfo {author} {\bibfnamefont {P.}~\bibnamefont {Coleman}}, \bibinfo {author} {\bibfnamefont {C.}~\bibnamefont {Geibel}}, \bibinfo {author} {\bibfnamefont {A.}~\bibnamefont {Steppke}}, \bibinfo {author} {\bibfnamefont {K.}~\bibnamefont {Kliemt}}, \bibinfo {author} {\bibfnamefont {C.}~\bibnamefont {Krellner}}, \bibinfo {author} {\bibfnamefont {M.}~\bibnamefont {Brando}},\ and\ \bibinfo {author} {\bibfnamefont {J.}~\bibnamefont
  {Saunders}},\ }\bibfield  {title} {\bibinfo {title} {Odd-parity superconductivity underpinned by antiferromagnetism in heavy fermion metal {YbRh$_2$Si$_2$}}\ }\href {https://doi.org/10.48550/ARXIV.2502.06420} {10.48550/ARXIV.2502.06420} (\bibinfo {year} {2025}),\ \Eprint {https://arxiv.org/abs/2502.06420} {arXiv:2502.06420 [cond-mat.supr-con]} \BibitemShut {NoStop}%
\bibitem [{\citenamefont {Knapp}\ \emph {et~al.}(2023)\citenamefont {Knapp}, \citenamefont {Levitin}, \citenamefont {Ny{\'{e}}ki}, \citenamefont {Ho}, \citenamefont {Cowan}, \citenamefont {Saunders}, \citenamefont {Brando}, \citenamefont {Geibel}, \citenamefont {Kliemt},\ and\ \citenamefont {Krellner}}]{Knapp2023}%
  \BibitemOpen
  \bibfield  {author} {\bibinfo {author} {\bibfnamefont {J.}~\bibnamefont {Knapp}}, \bibinfo {author} {\bibfnamefont {L.}~\bibnamefont {Levitin}}, \bibinfo {author} {\bibfnamefont {J.}~\bibnamefont {Ny{\'{e}}ki}}, \bibinfo {author} {\bibfnamefont {A.}~\bibnamefont {Ho}}, \bibinfo {author} {\bibfnamefont {B.}~\bibnamefont {Cowan}}, \bibinfo {author} {\bibfnamefont {J.}~\bibnamefont {Saunders}}, \bibinfo {author} {\bibfnamefont {M.}~\bibnamefont {Brando}}, \bibinfo {author} {\bibfnamefont {C.}~\bibnamefont {Geibel}}, \bibinfo {author} {\bibfnamefont {K.}~\bibnamefont {Kliemt}},\ and\ \bibinfo {author} {\bibfnamefont {C.}~\bibnamefont {Krellner}},\ }\bibfield  {title} {\bibinfo {title} {Electronuclear transition into a spatially modulated magnetic state in {YbRh$_2$Si$_2$}},\ }\href {https://doi.org/10.1103/physrevlett.130.126802} {\bibfield  {journal} {\bibinfo  {journal} {Physical Review Letters}\ }\textbf {\bibinfo {volume} {130}},\ \bibinfo {pages} {126802} (\bibinfo {year} {2023})}\BibitemShut {NoStop}%
\bibitem [{\citenamefont {Knapp}\ \emph {et~al.}(2024)\citenamefont {Knapp}, \citenamefont {Levitin}, \citenamefont {Nyéki}, \citenamefont {Brando},\ and\ \citenamefont {Saunders}}]{Knapp2024}%
  \BibitemOpen
  \bibfield  {author} {\bibinfo {author} {\bibfnamefont {J.}~\bibnamefont {Knapp}}, \bibinfo {author} {\bibfnamefont {L.~V.}\ \bibnamefont {Levitin}}, \bibinfo {author} {\bibfnamefont {J.}~\bibnamefont {Nyéki}}, \bibinfo {author} {\bibfnamefont {M.}~\bibnamefont {Brando}},\ and\ \bibinfo {author} {\bibfnamefont {J.}~\bibnamefont {Saunders}},\ }\bibfield  {title} {\bibinfo {title} {Precise calorimetry of small metal samples using noise thermometry},\ }\href {https://doi.org/10.1007/s10909-024-03207-w} {\bibfield  {journal} {\bibinfo  {journal} {Journal of Low Temperature Physics}\ }\textbf {\bibinfo {volume} {217}},\ \bibinfo {pages} {638} (\bibinfo {year} {2024})}\BibitemShut {NoStop}%
\bibitem [{\citenamefont {Friedemann}\ \emph {et~al.}(2010)\citenamefont {Friedemann}, \citenamefont {Wirth}, \citenamefont {Oeschler}, \citenamefont {Krellner}, \citenamefont {Geibel}, \citenamefont {Steglich}, \citenamefont {MaQuilon}, \citenamefont {Fisk}, \citenamefont {Paschen},\ and\ \citenamefont {Zwicknagl}}]{Friedemann2010b}%
  \BibitemOpen
  \bibfield  {author} {\bibinfo {author} {\bibfnamefont {S.}~\bibnamefont {Friedemann}}, \bibinfo {author} {\bibfnamefont {S.}~\bibnamefont {Wirth}}, \bibinfo {author} {\bibfnamefont {N.}~\bibnamefont {Oeschler}}, \bibinfo {author} {\bibfnamefont {C.}~\bibnamefont {Krellner}}, \bibinfo {author} {\bibfnamefont {C.}~\bibnamefont {Geibel}}, \bibinfo {author} {\bibfnamefont {F.}~\bibnamefont {Steglich}}, \bibinfo {author} {\bibfnamefont {S.}~\bibnamefont {MaQuilon}}, \bibinfo {author} {\bibfnamefont {Z.}~\bibnamefont {Fisk}}, \bibinfo {author} {\bibfnamefont {S.}~\bibnamefont {Paschen}},\ and\ \bibinfo {author} {\bibfnamefont {G.}~\bibnamefont {Zwicknagl}},\ }\bibfield  {title} {\bibinfo {title} {Hall effect measurements and electronic structure calculations on {YbRh$_2$Si$_2$} and its reference compounds {LuRh$_2$Si$_2$} and {YbIr$_2$Si$_2$}},\ }\href {https://doi.org/10.1103/physrevb.82.035103} {\bibfield  {journal} {\bibinfo  {journal} {Physical Review B}\ }\textbf {\bibinfo {volume} {82}},\ \bibinfo {pages}
  {035103} (\bibinfo {year} {2010})}\BibitemShut {NoStop}%
\bibitem [{\citenamefont {Kummer}\ \emph {et~al.}(2015)\citenamefont {Kummer}, \citenamefont {Patil}, \citenamefont {Chikina}, \citenamefont {G{\"{u}}ttler}, \citenamefont {H{\"{o}}ppner}, \citenamefont {Generalov}, \citenamefont {Danzenb{\"{a}}cher}, \citenamefont {Seiro}, \citenamefont {Hannaske}, \citenamefont {Krellner}, \citenamefont {Kucherenko}, \citenamefont {Shi}, \citenamefont {Radovic}, \citenamefont {Rienks}, \citenamefont {Zwicknagl}, \citenamefont {Matho}, \citenamefont {Allen}, \citenamefont {Laubschat}, \citenamefont {Geibel},\ and\ \citenamefont {Vyalikh}}]{Kummer2015}%
  \BibitemOpen
  \bibfield  {author} {\bibinfo {author} {\bibfnamefont {K.}~\bibnamefont {Kummer}}, \bibinfo {author} {\bibfnamefont {S.}~\bibnamefont {Patil}}, \bibinfo {author} {\bibfnamefont {A.}~\bibnamefont {Chikina}}, \bibinfo {author} {\bibfnamefont {M.}~\bibnamefont {G{\"{u}}ttler}}, \bibinfo {author} {\bibfnamefont {M.}~\bibnamefont {H{\"{o}}ppner}}, \bibinfo {author} {\bibfnamefont {A.}~\bibnamefont {Generalov}}, \bibinfo {author} {\bibfnamefont {S.}~\bibnamefont {Danzenb{\"{a}}cher}}, \bibinfo {author} {\bibfnamefont {S.}~\bibnamefont {Seiro}}, \bibinfo {author} {\bibfnamefont {A.}~\bibnamefont {Hannaske}}, \bibinfo {author} {\bibfnamefont {C.}~\bibnamefont {Krellner}}, \bibinfo {author} {\bibfnamefont {Y.}~\bibnamefont {Kucherenko}}, \bibinfo {author} {\bibfnamefont {M.}~\bibnamefont {Shi}}, \bibinfo {author} {\bibfnamefont {M.}~\bibnamefont {Radovic}}, \bibinfo {author} {\bibfnamefont {E.}~\bibnamefont {Rienks}}, \bibinfo {author} {\bibfnamefont {G.}~\bibnamefont {Zwicknagl}}, \bibinfo {author} {\bibfnamefont
  {K.}~\bibnamefont {Matho}}, \bibinfo {author} {\bibfnamefont {J.~W.}\ \bibnamefont {Allen}}, \bibinfo {author} {\bibfnamefont {C.}~\bibnamefont {Laubschat}}, \bibinfo {author} {\bibfnamefont {C.}~\bibnamefont {Geibel}},\ and\ \bibinfo {author} {\bibfnamefont {D.~V.}\ \bibnamefont {Vyalikh}},\ }\bibfield  {title} {\bibinfo {title} {Temperature-independent {Fermi} surface in the {Kondo} lattice {YbRh\textsubscript{2}Si\textsubscript{2}}},\ }\href {https://doi.org/10.1103/physrevx.5.011028} {\bibfield  {journal} {\bibinfo  {journal} {Physical Review X}\ }\textbf {\bibinfo {volume} {5}},\ \bibinfo {pages} {011028} (\bibinfo {year} {2015})}\BibitemShut {NoStop}%
\bibitem [{\citenamefont {Zwicknagl}(2016)}]{Zwicknagl2016}%
  \BibitemOpen
  \bibfield  {author} {\bibinfo {author} {\bibfnamefont {G.}~\bibnamefont {Zwicknagl}},\ }\bibfield  {title} {\bibinfo {title} {The utility of band theory in strongly correlated electron systems},\ }\href {https://doi.org/10.1088/0034-4885/79/12/124501} {\bibfield  {journal} {\bibinfo  {journal} {Reports on Progress in Physics}\ }\textbf {\bibinfo {volume} {79}},\ \bibinfo {pages} {124501} (\bibinfo {year} {2016})}\BibitemShut {NoStop}%
\bibitem [{\citenamefont {Li}\ \emph {et~al.}(2019)\citenamefont {Li}, \citenamefont {Wang}, \citenamefont {Xu}, \citenamefont {Xie},\ and\ \citenamefont {Yang}}]{Li2019a}%
  \BibitemOpen
  \bibfield  {author} {\bibinfo {author} {\bibfnamefont {Y.}~\bibnamefont {Li}}, \bibinfo {author} {\bibfnamefont {Q.}~\bibnamefont {Wang}}, \bibinfo {author} {\bibfnamefont {Y.}~\bibnamefont {Xu}}, \bibinfo {author} {\bibfnamefont {W.}~\bibnamefont {Xie}},\ and\ \bibinfo {author} {\bibfnamefont {Y.~F.}\ \bibnamefont {Yang}},\ }\bibfield  {title} {\bibinfo {title} {Nearly degenerate $p_x+ip_y$ and $d_{x^2-y^2}$ pairing symmetry in the heavy fermion superconductor {YbRh$_2$Si$_2$}},\ }\href {https://doi.org/10.1103/physrevb.100.085132} {\bibfield  {journal} {\bibinfo  {journal} {Physical Review B}\ }\textbf {\bibinfo {volume} {100}},\ \bibinfo {pages} {085132} (\bibinfo {year} {2019})}\BibitemShut {NoStop}%
\bibitem [{\citenamefont {G{\"{u}}ttler}\ \emph {et~al.}(2021)\citenamefont {G{\"{u}}ttler}, \citenamefont {Kummer}, \citenamefont {Kliemt}, \citenamefont {Krellner}, \citenamefont {Seiro}, \citenamefont {Geibel}, \citenamefont {Laubschat}, \citenamefont {Kubo}, \citenamefont {Sakurai}, \citenamefont {Vyalikh},\ and\ \citenamefont {Koizumi}}]{Guettler2021}%
  \BibitemOpen
  \bibfield  {author} {\bibinfo {author} {\bibfnamefont {M.}~\bibnamefont {G{\"{u}}ttler}}, \bibinfo {author} {\bibfnamefont {K.}~\bibnamefont {Kummer}}, \bibinfo {author} {\bibfnamefont {K.}~\bibnamefont {Kliemt}}, \bibinfo {author} {\bibfnamefont {C.}~\bibnamefont {Krellner}}, \bibinfo {author} {\bibfnamefont {S.}~\bibnamefont {Seiro}}, \bibinfo {author} {\bibfnamefont {C.}~\bibnamefont {Geibel}}, \bibinfo {author} {\bibfnamefont {C.}~\bibnamefont {Laubschat}}, \bibinfo {author} {\bibfnamefont {Y.}~\bibnamefont {Kubo}}, \bibinfo {author} {\bibfnamefont {Y.}~\bibnamefont {Sakurai}}, \bibinfo {author} {\bibfnamefont {D.~V.}\ \bibnamefont {Vyalikh}},\ and\ \bibinfo {author} {\bibfnamefont {A.}~\bibnamefont {Koizumi}},\ }\bibfield  {title} {\bibinfo {title} {Visualizing the {Kondo} lattice crossover in {YbRh\textsubscript{2}Si\textsubscript{2}} with {Compton} scattering},\ }\href {https://doi.org/10.1103/physrevb.103.115126} {\bibfield  {journal} {\bibinfo  {journal} {Physical Review B}\ }\textbf {\bibinfo
  {volume} {103}},\ \bibinfo {pages} {115126} (\bibinfo {year} {2021})}\BibitemShut {NoStop}%
\bibitem [{\citenamefont {Ishida}\ \emph {et~al.}(2003)\citenamefont {Ishida}, \citenamefont {MacLaughlin}, \citenamefont {Young}, \citenamefont {Okamoto}, \citenamefont {Kawasaki}, \citenamefont {Kitaoka}, \citenamefont {Nieuwenhuys}, \citenamefont {Heffner}, \citenamefont {Bernal}, \citenamefont {Higemoto}, \citenamefont {Koda}, \citenamefont {Kadono}, \citenamefont {Trovarelli}, \citenamefont {Geibel},\ and\ \citenamefont {Steglich}}]{Ishida2003}%
  \BibitemOpen
  \bibfield  {author} {\bibinfo {author} {\bibfnamefont {K.}~\bibnamefont {Ishida}}, \bibinfo {author} {\bibfnamefont {D.~E.}\ \bibnamefont {MacLaughlin}}, \bibinfo {author} {\bibfnamefont {B.~L.}\ \bibnamefont {Young}}, \bibinfo {author} {\bibfnamefont {K.}~\bibnamefont {Okamoto}}, \bibinfo {author} {\bibfnamefont {Y.}~\bibnamefont {Kawasaki}}, \bibinfo {author} {\bibfnamefont {Y.}~\bibnamefont {Kitaoka}}, \bibinfo {author} {\bibfnamefont {G.~J.}\ \bibnamefont {Nieuwenhuys}}, \bibinfo {author} {\bibfnamefont {R.~H.}\ \bibnamefont {Heffner}}, \bibinfo {author} {\bibfnamefont {O.~O.}\ \bibnamefont {Bernal}}, \bibinfo {author} {\bibfnamefont {W.}~\bibnamefont {Higemoto}}, \bibinfo {author} {\bibfnamefont {A.}~\bibnamefont {Koda}}, \bibinfo {author} {\bibfnamefont {R.}~\bibnamefont {Kadono}}, \bibinfo {author} {\bibfnamefont {O.}~\bibnamefont {Trovarelli}}, \bibinfo {author} {\bibfnamefont {C.}~\bibnamefont {Geibel}},\ and\ \bibinfo {author} {\bibfnamefont {F.}~\bibnamefont {Steglich}},\ }\bibfield  {title}
  {\bibinfo {title} {Low{\textendash}temperature magnetic order and spin dynamics in {YbRh}$_2${Si}$_2$},\ }\href {https://doi.org/10.1103/physrevb.68.184401} {\bibfield  {journal} {\bibinfo  {journal} {Physical Review B}\ }\textbf {\bibinfo {volume} {68}},\ \bibinfo {pages} {184401} (\bibinfo {year} {2003})}\BibitemShut {NoStop}%
\bibitem [{\citenamefont {Hamann}\ \emph {et~al.}(2019)\citenamefont {Hamann}, \citenamefont {Zhang}, \citenamefont {Jang}, \citenamefont {Hannaske}, \citenamefont {Steinke}, \citenamefont {Lausberg}, \citenamefont {Pedrero}, \citenamefont {Klingner}, \citenamefont {Baenitz}, \citenamefont {Steglich}, \citenamefont {Krellner}, \citenamefont {Geibel},\ and\ \citenamefont {Brando}}]{Hamann2019}%
  \BibitemOpen
  \bibfield  {author} {\bibinfo {author} {\bibfnamefont {S.}~\bibnamefont {Hamann}}, \bibinfo {author} {\bibfnamefont {J.}~\bibnamefont {Zhang}}, \bibinfo {author} {\bibfnamefont {D.}~\bibnamefont {Jang}}, \bibinfo {author} {\bibfnamefont {A.}~\bibnamefont {Hannaske}}, \bibinfo {author} {\bibfnamefont {L.}~\bibnamefont {Steinke}}, \bibinfo {author} {\bibfnamefont {S.}~\bibnamefont {Lausberg}}, \bibinfo {author} {\bibfnamefont {L.}~\bibnamefont {Pedrero}}, \bibinfo {author} {\bibfnamefont {C.}~\bibnamefont {Klingner}}, \bibinfo {author} {\bibfnamefont {M.}~\bibnamefont {Baenitz}}, \bibinfo {author} {\bibfnamefont {F.}~\bibnamefont {Steglich}}, \bibinfo {author} {\bibfnamefont {C.}~\bibnamefont {Krellner}}, \bibinfo {author} {\bibfnamefont {C.}~\bibnamefont {Geibel}},\ and\ \bibinfo {author} {\bibfnamefont {M.}~\bibnamefont {Brando}},\ }\bibfield  {title} {\bibinfo {title} {Evolution from ferromagnetism to antiferromagnetism in {Yb(Rh\textsubscript{1-x}Co\textsubscript{x})Si\textsubscript{2}}},\ }\href
  {https://doi.org/10.1103/physrevlett.122.077202} {\bibfield  {journal} {\bibinfo  {journal} {Physical Review Letters}\ }\textbf {\bibinfo {volume} {122}},\ \bibinfo {pages} {077202} (\bibinfo {year} {2019})}\BibitemShut {NoStop}%
\bibitem [{\citenamefont {Si}\ \emph {et~al.}(2001)\citenamefont {Si}, \citenamefont {Rabello}, \citenamefont {Ingersent},\ and\ \citenamefont {Smith}}]{Si2001}%
  \BibitemOpen
  \bibfield  {author} {\bibinfo {author} {\bibfnamefont {Q.}~\bibnamefont {Si}}, \bibinfo {author} {\bibfnamefont {S.}~\bibnamefont {Rabello}}, \bibinfo {author} {\bibfnamefont {K.}~\bibnamefont {Ingersent}},\ and\ \bibinfo {author} {\bibfnamefont {J.~L.}\ \bibnamefont {Smith}},\ }\bibfield  {title} {\bibinfo {title} {Locally critical quantum phase transitions in strongly correlated metals},\ }\href {https://doi.org/10.1038/35101507} {\bibfield  {journal} {\bibinfo  {journal} {Nature}\ }\textbf {\bibinfo {volume} {413}},\ \bibinfo {pages} {804} (\bibinfo {year} {2001})}\BibitemShut {NoStop}%
\bibitem [{\citenamefont {Coleman}\ and\ \citenamefont {Schofield}(2005)}]{Coleman2005}%
  \BibitemOpen
  \bibfield  {author} {\bibinfo {author} {\bibfnamefont {P.}~\bibnamefont {Coleman}}\ and\ \bibinfo {author} {\bibfnamefont {A.~J.}\ \bibnamefont {Schofield}},\ }\bibfield  {title} {\bibinfo {title} {Quantum criticality},\ }\href {https://doi.org/10.1038/nature03279} {\bibfield  {journal} {\bibinfo  {journal} {Nature}\ }\textbf {\bibinfo {volume} {433}},\ \bibinfo {pages} {226} (\bibinfo {year} {2005})}\BibitemShut {NoStop}%
\bibitem [{\citenamefont {Si}\ and\ \citenamefont {Steglich}(2010)}]{Si2010}%
  \BibitemOpen
  \bibfield  {author} {\bibinfo {author} {\bibfnamefont {Q.}~\bibnamefont {Si}}\ and\ \bibinfo {author} {\bibfnamefont {F.}~\bibnamefont {Steglich}},\ }\bibfield  {title} {\bibinfo {title} {Heavy {Fermions} and quantum phase transitions},\ }\href {https://doi.org/10.1126/science.1191195} {\bibfield  {journal} {\bibinfo  {journal} {Science}\ }\textbf {\bibinfo {volume} {329}},\ \bibinfo {pages} {1161} (\bibinfo {year} {2010})}\BibitemShut {NoStop}%
\bibitem [{\citenamefont {Steglich}(2014)}]{Steglich2014}%
  \BibitemOpen
  \bibfield  {author} {\bibinfo {author} {\bibfnamefont {F.}~\bibnamefont {Steglich}},\ }\bibfield  {title} {\bibinfo {title} {Heavy fermions: superconductivity and its relationship to quantum criticality},\ }\href {https://doi.org/10.1080/14786435.2014.956835} {\bibfield  {journal} {\bibinfo  {journal} {Philosophical Magazine}\ }\textbf {\bibinfo {volume} {94}},\ \bibinfo {pages} {3259} (\bibinfo {year} {2014})}\BibitemShut {NoStop}%
\bibitem [{\citenamefont {Schubert}\ \emph {et~al.}(2019)\citenamefont {Schubert}, \citenamefont {Tokiwa}, \citenamefont {H{\"{u}}bner}, \citenamefont {Mchalwat}, \citenamefont {Blumenr{\"{o}}ther}, \citenamefont {Jeevan},\ and\ \citenamefont {Gegenwart}}]{Schubert2019}%
  \BibitemOpen
  \bibfield  {author} {\bibinfo {author} {\bibfnamefont {M.~H.}\ \bibnamefont {Schubert}}, \bibinfo {author} {\bibfnamefont {Y.}~\bibnamefont {Tokiwa}}, \bibinfo {author} {\bibfnamefont {S.~H.}\ \bibnamefont {H{\"{u}}bner}}, \bibinfo {author} {\bibfnamefont {M.}~\bibnamefont {Mchalwat}}, \bibinfo {author} {\bibfnamefont {E.}~\bibnamefont {Blumenr{\"{o}}ther}}, \bibinfo {author} {\bibfnamefont {H.~S.}\ \bibnamefont {Jeevan}},\ and\ \bibinfo {author} {\bibfnamefont {P.}~\bibnamefont {Gegenwart}},\ }\bibfield  {title} {\bibinfo {title} {Tuning low-energy scales in {YbRh$_2$Si$_2$} by non-isoelectronic substitution and pressure},\ }\href {https://doi.org/10.1103/physrevresearch.1.032004} {\bibfield  {journal} {\bibinfo  {journal} {Physical Review Research}\ }\textbf {\bibinfo {volume} {1}},\ \bibinfo {pages} {032004(R)} (\bibinfo {year} {2019})}\BibitemShut {NoStop}%
\bibitem [{\citenamefont {Paschen}\ and\ \citenamefont {Si}(2020)}]{Paschen2020}%
  \BibitemOpen
  \bibfield  {author} {\bibinfo {author} {\bibfnamefont {S.}~\bibnamefont {Paschen}}\ and\ \bibinfo {author} {\bibfnamefont {Q.}~\bibnamefont {Si}},\ }\bibfield  {title} {\bibinfo {title} {Quantum phases driven by strong correlations},\ }\href {https://doi.org/10.1038/s42254-020-00262-6} {\bibfield  {journal} {\bibinfo  {journal} {Nature Reviews Physics}\ }\textbf {\bibinfo {volume} {3}},\ \bibinfo {pages} {9} (\bibinfo {year} {2020})}\BibitemShut {NoStop}%
\bibitem [{\citenamefont {Schuberth}\ \emph {et~al.}(2022)\citenamefont {Schuberth}, \citenamefont {Wirth},\ and\ \citenamefont {Steglich}}]{Schuberth2022}%
  \BibitemOpen
  \bibfield  {author} {\bibinfo {author} {\bibfnamefont {E.}~\bibnamefont {Schuberth}}, \bibinfo {author} {\bibfnamefont {S.}~\bibnamefont {Wirth}},\ and\ \bibinfo {author} {\bibfnamefont {F.}~\bibnamefont {Steglich}},\ }\bibfield  {title} {\bibinfo {title} {Nuclear-order-induced quantum criticality and {Heavy-Fermion} superconductivity at ultra-low temperatures in {YbRh}$_2${Si}$_2$},\ }\bibfield  {journal} {\bibinfo  {journal} {Frontiers in Electronic Materials}\ }\textbf {\bibinfo {volume} {2}},\ \href {https://doi.org/10.3389/femat.2022.869495} {10.3389/femat.2022.869495} (\bibinfo {year} {2022})\BibitemShut {NoStop}%
\bibitem [{\citenamefont {Abrahams}\ and\ \citenamefont {W{\"{o}}lfle}(2012)}]{Abrahams2012}%
  \BibitemOpen
  \bibfield  {author} {\bibinfo {author} {\bibfnamefont {E.}~\bibnamefont {Abrahams}}\ and\ \bibinfo {author} {\bibfnamefont {P.}~\bibnamefont {W{\"{o}}lfle}},\ }\bibfield  {title} {\bibinfo {title} {Critical quasiparticle theory applied to heavy fermion metals near an antiferromagnetic quantum phase transition},\ }\href {https://doi.org/10.1073/pnas.1200346109} {\bibfield  {journal} {\bibinfo  {journal} {Proceedings of the National Academy of Sciences}\ }\textbf {\bibinfo {volume} {109}},\ \bibinfo {pages} {3238} (\bibinfo {year} {2012})}\BibitemShut {NoStop}%
\bibitem [{\citenamefont {Abrahams}\ \emph {et~al.}(2014)\citenamefont {Abrahams}, \citenamefont {Schmalian},\ and\ \citenamefont {W{\"{o}}lfle}}]{Abrahams2014}%
  \BibitemOpen
  \bibfield  {author} {\bibinfo {author} {\bibfnamefont {E.}~\bibnamefont {Abrahams}}, \bibinfo {author} {\bibfnamefont {J.}~\bibnamefont {Schmalian}},\ and\ \bibinfo {author} {\bibfnamefont {P.}~\bibnamefont {W{\"{o}}lfle}},\ }\bibfield  {title} {\bibinfo {title} {Strong-coupling theory of heavy-fermion criticality},\ }\href {https://doi.org/10.1103/physrevb.90.045105} {\bibfield  {journal} {\bibinfo  {journal} {Physical Review B}\ }\textbf {\bibinfo {volume} {90}},\ \bibinfo {pages} {045105} (\bibinfo {year} {2014})}\BibitemShut {NoStop}%
\bibitem [{\citenamefont {W{\"{o}}lfle}\ and\ \citenamefont {Abrahams}(2015)}]{Woelfle2015}%
  \BibitemOpen
  \bibfield  {author} {\bibinfo {author} {\bibfnamefont {P.}~\bibnamefont {W{\"{o}}lfle}}\ and\ \bibinfo {author} {\bibfnamefont {E.}~\bibnamefont {Abrahams}},\ }\bibfield  {title} {\bibinfo {title} {Spin-flip scattering of critical quasiparticles and the phase diagram of {YbRh\textsubscript{2}Si\textsubscript{2}}},\ }\href {https://doi.org/10.1103/physrevb.92.155111} {\bibfield  {journal} {\bibinfo  {journal} {Physical Review B}\ }\textbf {\bibinfo {volume} {92}},\ \bibinfo {pages} {155111} (\bibinfo {year} {2015})}\BibitemShut {NoStop}%
\bibitem [{\citenamefont {Watanabe}\ and\ \citenamefont {Miyake}(2010)}]{Watanabe2010}%
  \BibitemOpen
  \bibfield  {author} {\bibinfo {author} {\bibfnamefont {S.}~\bibnamefont {Watanabe}}\ and\ \bibinfo {author} {\bibfnamefont {K.}~\bibnamefont {Miyake}},\ }\bibfield  {title} {\bibinfo {title} {Quantum valence criticality as an origin of unconventional critical phenomena},\ }\href {https://doi.org/10.1103/physrevlett.105.186403} {\bibfield  {journal} {\bibinfo  {journal} {Physical Review Letters}\ }\textbf {\bibinfo {volume} {105}},\ \bibinfo {pages} {186403} (\bibinfo {year} {2010})}\BibitemShut {NoStop}%
\bibitem [{\citenamefont {Steinke}\ \emph {et~al.}(2017)\citenamefont {Steinke}, \citenamefont {Schuberth}, \citenamefont {Lausberg}, \citenamefont {Tippmann}, \citenamefont {Steppke}, \citenamefont {Krellner}, \citenamefont {Geibel}, \citenamefont {Steglich},\ and\ \citenamefont {Brando}}]{Steinke2017}%
  \BibitemOpen
  \bibfield  {author} {\bibinfo {author} {\bibfnamefont {L.}~\bibnamefont {Steinke}}, \bibinfo {author} {\bibfnamefont {E.}~\bibnamefont {Schuberth}}, \bibinfo {author} {\bibfnamefont {S.}~\bibnamefont {Lausberg}}, \bibinfo {author} {\bibfnamefont {M.}~\bibnamefont {Tippmann}}, \bibinfo {author} {\bibfnamefont {A.}~\bibnamefont {Steppke}}, \bibinfo {author} {\bibfnamefont {C.}~\bibnamefont {Krellner}}, \bibinfo {author} {\bibfnamefont {C.}~\bibnamefont {Geibel}}, \bibinfo {author} {\bibfnamefont {F.}~\bibnamefont {Steglich}},\ and\ \bibinfo {author} {\bibfnamefont {M.}~\bibnamefont {Brando}},\ }\bibfield  {title} {\bibinfo {title} {Ultra{\textendash}low temperature ac susceptibility of the heavy{\textendash}fermion superconductor {YbRh$_2$Si$_2$}},\ }\href {https://doi.org/10.1088/1742-6596/807/5/052007} {\bibfield  {journal} {\bibinfo  {journal} {Journal of Physics: Conference Series}\ }\textbf {\bibinfo {volume} {807}},\ \bibinfo {pages} {052007} (\bibinfo {year} {2017})}\BibitemShut {NoStop}%
\bibitem [{SM()}]{SM}%
  \BibitemOpen
  \href@noop {} {}\bibinfo {note} {See Supplemental Material at [URL will be inserted by the publisher], including an additional reference \cite{Andres1982}, for details of the hyperfine fields calculations, the hyperfine Schottky model, the thermodynamics of the magneto-caloric sweeps, the phase boudaries and critical exponents, and the investigation of hysteresis and preparation condition dependence.}\BibitemShut {Stop}%
\bibitem [{\citenamefont {GschneidnerJr}\ \emph {et~al.}(2005)\citenamefont {GschneidnerJr}, \citenamefont {Pecharsky},\ and\ \citenamefont {Tsokol}}]{GschneidnerJr2005}%
  \BibitemOpen
  \bibfield  {author} {\bibinfo {author} {\bibfnamefont {K.~A.}\ \bibnamefont {GschneidnerJr}}, \bibinfo {author} {\bibfnamefont {V.~K.}\ \bibnamefont {Pecharsky}},\ and\ \bibinfo {author} {\bibfnamefont {A.~O.}\ \bibnamefont {Tsokol}},\ }\bibfield  {title} {\bibinfo {title} {Recent developments in magnetocaloric materials},\ }\href {https://doi.org/10.1088/0034-4885/68/6/r04} {\bibfield  {journal} {\bibinfo  {journal} {Reports on Progress in Physics}\ }\textbf {\bibinfo {volume} {68}},\ \bibinfo {pages} {1479} (\bibinfo {year} {2005})}\BibitemShut {NoStop}%
\bibitem [{\citenamefont {Pecharsky}\ \emph {et~al.}(2001)\citenamefont {Pecharsky}, \citenamefont {Gschneidner}, \citenamefont {Pecharsky},\ and\ \citenamefont {Tishin}}]{Pecharsky2001}%
  \BibitemOpen
  \bibfield  {author} {\bibinfo {author} {\bibfnamefont {V.}~\bibnamefont {Pecharsky}}, \bibinfo {author} {\bibfnamefont {K.}~\bibnamefont {Gschneidner}}, \bibinfo {author} {\bibfnamefont {A.}~\bibnamefont {Pecharsky}},\ and\ \bibinfo {author} {\bibfnamefont {A.}~\bibnamefont {Tishin}},\ }\bibfield  {title} {\bibinfo {title} {Thermodynamics of the magnetocaloric effect},\ }\href {https://doi.org/10.1103/physrevb.64.144406} {\bibfield  {journal} {\bibinfo  {journal} {Physical Review B}\ }\textbf {\bibinfo {volume} {64}},\ \bibinfo {pages} {144406} (\bibinfo {year} {2001})}\BibitemShut {NoStop}%
\bibitem [{\citenamefont {Pecharsky}\ and\ \citenamefont {Gschneidner~Jr.}(1997)}]{Pecharsky1997}%
  \BibitemOpen
  \bibfield  {author} {\bibinfo {author} {\bibfnamefont {V.~K.}\ \bibnamefont {Pecharsky}}\ and\ \bibinfo {author} {\bibfnamefont {K.~A.}\ \bibnamefont {Gschneidner~Jr.}},\ }\bibfield  {title} {\bibinfo {title} {Giant {{Magnetocaloric Effect}} in {{Gd}}$_5$({{Si}}$_2${{Ge}}$_2$)},\ }\href {https://doi.org/10.1103/PhysRevLett.78.4494} {\bibfield  {journal} {\bibinfo  {journal} {Physical Review Letters}\ }\textbf {\bibinfo {volume} {78}},\ \bibinfo {pages} {4494} (\bibinfo {year} {1997})}\BibitemShut {NoStop}%
\bibitem [{\citenamefont {Krellner}\ \emph {et~al.}(2009)\citenamefont {Krellner}, \citenamefont {Hartmann}, \citenamefont {Pikul}, \citenamefont {Oeschler}, \citenamefont {Donath}, \citenamefont {Geibel}, \citenamefont {Steglich},\ and\ \citenamefont {Wosnitza}}]{Krellner2009}%
  \BibitemOpen
  \bibfield  {author} {\bibinfo {author} {\bibfnamefont {C.}~\bibnamefont {Krellner}}, \bibinfo {author} {\bibfnamefont {S.}~\bibnamefont {Hartmann}}, \bibinfo {author} {\bibfnamefont {A.}~\bibnamefont {Pikul}}, \bibinfo {author} {\bibfnamefont {N.}~\bibnamefont {Oeschler}}, \bibinfo {author} {\bibfnamefont {J.~G.}\ \bibnamefont {Donath}}, \bibinfo {author} {\bibfnamefont {C.}~\bibnamefont {Geibel}}, \bibinfo {author} {\bibfnamefont {F.}~\bibnamefont {Steglich}},\ and\ \bibinfo {author} {\bibfnamefont {J.}~\bibnamefont {Wosnitza}},\ }\bibfield  {title} {\bibinfo {title} {Violation of critical universality at the antiferromagnetic phase transition of {YbRh}$_2${Si}$_2$},\ }\href {https://doi.org/10.1103/physrevlett.102.196402} {\bibfield  {journal} {\bibinfo  {journal} {Physical Review Letters}\ }\textbf {\bibinfo {volume} {102}},\ \bibinfo {pages} {196402} (\bibinfo {year} {2009})}\BibitemShut {NoStop}%
\bibitem [{\citenamefont {Nowik}\ and\ \citenamefont {Ofer}(1968)}]{Nowik1968}%
  \BibitemOpen
  \bibfield  {author} {\bibinfo {author} {\bibfnamefont {I.}~\bibnamefont {Nowik}}\ and\ \bibinfo {author} {\bibfnamefont {S.}~\bibnamefont {Ofer}},\ }\bibfield  {title} {\bibinfo {title} {{Mössbauer} studies of $^{170}${Yb} in several paramagnetic salts},\ }\href {https://doi.org/10.1016/0022-3697(68)90007-3} {\bibfield  {journal} {\bibinfo  {journal} {Journal of Physics and Chemistry of Solids}\ }\textbf {\bibinfo {volume} {29}},\ \bibinfo {pages} {2117} (\bibinfo {year} {1968})}\BibitemShut {NoStop}%
\bibitem [{\citenamefont {Plessel}\ \emph {et~al.}(2003)\citenamefont {Plessel}, \citenamefont {Abd{\textendash}Elmeguid}, \citenamefont {Sanchez}, \citenamefont {Knebel}, \citenamefont {Geibel}, \citenamefont {Trovarelli},\ and\ \citenamefont {Steglich}}]{Plessel2003}%
  \BibitemOpen
  \bibfield  {author} {\bibinfo {author} {\bibfnamefont {J.}~\bibnamefont {Plessel}}, \bibinfo {author} {\bibfnamefont {M.~M.}\ \bibnamefont {Abd{\textendash}Elmeguid}}, \bibinfo {author} {\bibfnamefont {J.~P.}\ \bibnamefont {Sanchez}}, \bibinfo {author} {\bibfnamefont {G.}~\bibnamefont {Knebel}}, \bibinfo {author} {\bibfnamefont {C.}~\bibnamefont {Geibel}}, \bibinfo {author} {\bibfnamefont {O.}~\bibnamefont {Trovarelli}},\ and\ \bibinfo {author} {\bibfnamefont {F.}~\bibnamefont {Steglich}},\ }\bibfield  {title} {\bibinfo {title} {Unusual behavior of the low{\textendash}moment magnetic ground state of {YbRh}$_2${Si}$_2$ under high pressure},\ }\href {https://doi.org/10.1103/physrevb.67.180403} {\bibfield  {journal} {\bibinfo  {journal} {Physical Review B}\ }\textbf {\bibinfo {volume} {67}},\ \bibinfo {pages} {180403(R)} (\bibinfo {year} {2003})}\BibitemShut {NoStop}%
\bibitem [{\citenamefont {Flouquet}\ and\ \citenamefont {Brewer}(1975)}]{Flouquet1975}%
  \BibitemOpen
  \bibfield  {author} {\bibinfo {author} {\bibfnamefont {J.}~\bibnamefont {Flouquet}}\ and\ \bibinfo {author} {\bibfnamefont {W.~D.}\ \bibnamefont {Brewer}},\ }\bibfield  {title} {\bibinfo {title} {Hyperfine interaction studies of local moments in metals},\ }\href {https://doi.org/10.1088/0031-8949/11/3-4/013} {\bibfield  {journal} {\bibinfo  {journal} {Physica Scripta}\ }\textbf {\bibinfo {volume} {11}},\ \bibinfo {pages} {199} (\bibinfo {year} {1975})}\BibitemShut {NoStop}%
\bibitem [{\citenamefont {Flouquet}(1978)}]{Flouquet1978}%
  \BibitemOpen
  \bibfield  {author} {\bibinfo {author} {\bibfnamefont {J.}~\bibnamefont {Flouquet}},\ }\bibfield  {title} {\bibinfo {title} {Kondo coupling, hyperfine and exchange interactions},\ }\href {https://doi.org/10.1051/jphyscol:19786592} {\bibfield  {journal} {\bibinfo  {journal} {Le Journal de Physique Colloques}\ }\textbf {\bibinfo {volume} {39}},\ \bibinfo {pages} {C6} (\bibinfo {year} {1978})}\BibitemShut {NoStop}%
\bibitem [{\citenamefont {Bonville}\ \emph {et~al.}(1992)\citenamefont {Bonville}, \citenamefont {Canaud}, \citenamefont {Hammann}, \citenamefont {Hodges}, \citenamefont {Imbert}, \citenamefont {Jéhanno}, \citenamefont {Severing},\ and\ \citenamefont {Fisk}}]{Bonville1992}%
  \BibitemOpen
  \bibfield  {author} {\bibinfo {author} {\bibfnamefont {P.}~\bibnamefont {Bonville}}, \bibinfo {author} {\bibfnamefont {B.}~\bibnamefont {Canaud}}, \bibinfo {author} {\bibfnamefont {J.}~\bibnamefont {Hammann}}, \bibinfo {author} {\bibfnamefont {J.~A.}\ \bibnamefont {Hodges}}, \bibinfo {author} {\bibfnamefont {P.}~\bibnamefont {Imbert}}, \bibinfo {author} {\bibfnamefont {G.}~\bibnamefont {Jéhanno}}, \bibinfo {author} {\bibfnamefont {A.}~\bibnamefont {Severing}},\ and\ \bibinfo {author} {\bibfnamefont {Z.}~\bibnamefont {Fisk}},\ }\bibfield  {title} {\bibinfo {title} {Magnetic ordering and hybridisation in ybaucu$_4$},\ }\href {https://doi.org/10.1051/jp1:1992157} {\bibfield  {journal} {\bibinfo  {journal} {Journal de Physique I}\ }\textbf {\bibinfo {volume} {2}},\ \bibinfo {pages} {459} (\bibinfo {year} {1992})}\BibitemShut {NoStop}%
\bibitem [{\citenamefont {Bonville}\ \emph {et~al.}(1991)\citenamefont {Bonville}, \citenamefont {Hodges}, \citenamefont {Imbert}, \citenamefont {J{\'{e}}hanno}, \citenamefont {Jaccard},\ and\ \citenamefont {Sierro}}]{Bonville1991}%
  \BibitemOpen
  \bibfield  {author} {\bibinfo {author} {\bibfnamefont {P.}~\bibnamefont {Bonville}}, \bibinfo {author} {\bibfnamefont {J.~A.}\ \bibnamefont {Hodges}}, \bibinfo {author} {\bibfnamefont {P.}~\bibnamefont {Imbert}}, \bibinfo {author} {\bibfnamefont {G.}~\bibnamefont {J{\'{e}}hanno}}, \bibinfo {author} {\bibfnamefont {D.}~\bibnamefont {Jaccard}},\ and\ \bibinfo {author} {\bibfnamefont {J.}~\bibnamefont {Sierro}},\ }\bibfield  {title} {\bibinfo {title} {Magnetic ordering and paramagnetic relaxation of {Yb}$^{3+}$ in {YbNi}$_2${Si}$_2$},\ }\href {https://doi.org/10.1016/0304-8853(91)90178-d} {\bibfield  {journal} {\bibinfo  {journal} {Journal of Magnetism and Magnetic Materials}\ }\textbf {\bibinfo {volume} {97}},\ \bibinfo {pages} {178} (\bibinfo {year} {1991})}\BibitemShut {NoStop}%
\bibitem [{\citenamefont {Bonville}\ \emph {et~al.}(1984)\citenamefont {Bonville}, \citenamefont {Imbert}, \citenamefont {J{\'{e}}hanno}, \citenamefont {Gonzalez{\textendash}Jimenez},\ and\ \citenamefont {Hartmann{\textendash}Boutron}}]{Bonville1984}%
  \BibitemOpen
  \bibfield  {author} {\bibinfo {author} {\bibfnamefont {P.}~\bibnamefont {Bonville}}, \bibinfo {author} {\bibfnamefont {P.}~\bibnamefont {Imbert}}, \bibinfo {author} {\bibfnamefont {G.}~\bibnamefont {J{\'{e}}hanno}}, \bibinfo {author} {\bibfnamefont {F.}~\bibnamefont {Gonzalez{\textendash}Jimenez}},\ and\ \bibinfo {author} {\bibfnamefont {F.}~\bibnamefont {Hartmann{\textendash}Boutron}},\ }\bibfield  {title} {\bibinfo {title} {Emission {M{\"{o}}ssbauer} spectroscopy and relaxation measurements in hyperfine levels out of thermal equilibrium: Very{\textendash}low{\textendash}temperature experiments on the {Kondo} {alloy Au$^{170}$Yb}},\ }\href {https://doi.org/10.1103/physrevb.30.3672} {\bibfield  {journal} {\bibinfo  {journal} {Physical Review B}\ }\textbf {\bibinfo {volume} {30}},\ \bibinfo {pages} {3672} (\bibinfo {year} {1984})}\BibitemShut {NoStop}%
\bibitem [{\citenamefont {Kondo}(1961)}]{Kondo1961}%
  \BibitemOpen
  \bibfield  {author} {\bibinfo {author} {\bibfnamefont {J.}~\bibnamefont {Kondo}},\ }\bibfield  {title} {\bibinfo {title} {Internal magnetic field in rare earth metals},\ }\href {https://doi.org/10.1143/jpsj.16.1690} {\bibfield  {journal} {\bibinfo  {journal} {Journal of the Physical Society of Japan}\ }\textbf {\bibinfo {volume} {16}},\ \bibinfo {pages} {1690} (\bibinfo {year} {1961})}\BibitemShut {NoStop}%
\bibitem [{Note1()}]{Note1}%
  \BibitemOpen
  \bibinfo {note} {The measured susceptibility leads to the enhancement of the magnetic induction $B=(1+\chi )B_{\protect \mathrm {ext}}$ inside YbRh\protect \textsubscript {2}Si\protect \textsubscript {2} over the external field $B_{\protect \mathrm {ext}}$ by about 20\%. This effect should be incorporated in future theoretical models of the magnetism in YbRh\protect \textsubscript {2}Si\protect \textsubscript {2}.}\BibitemShut {Stop}%
\bibitem [{\citenamefont {Continentino}(1998)}]{Continentino1998}%
  \BibitemOpen
  \bibfield  {author} {\bibinfo {author} {\bibfnamefont {M.~A.}\ \bibnamefont {Continentino}},\ }\bibfield  {title} {\bibinfo {title} {Universality in heavy fermions},\ }\href {https://doi.org/10.1103/physrevb.57.5966} {\bibfield  {journal} {\bibinfo  {journal} {Physical Review B}\ }\textbf {\bibinfo {volume} {57}},\ \bibinfo {pages} {5966} (\bibinfo {year} {1998})}\BibitemShut {NoStop}%
\bibitem [{\citenamefont {Chandra}\ \emph {et~al.}(2020)\citenamefont {Chandra}, \citenamefont {Coleman}, \citenamefont {Continentino},\ and\ \citenamefont {Lonzarich}}]{Chandra2020}%
  \BibitemOpen
  \bibfield  {author} {\bibinfo {author} {\bibfnamefont {P.}~\bibnamefont {Chandra}}, \bibinfo {author} {\bibfnamefont {P.}~\bibnamefont {Coleman}}, \bibinfo {author} {\bibfnamefont {M.~A.}\ \bibnamefont {Continentino}},\ and\ \bibinfo {author} {\bibfnamefont {G.~G.}\ \bibnamefont {Lonzarich}},\ }\bibfield  {title} {\bibinfo {title} {Quantum annealed criticality: A scaling description},\ }\href {https://doi.org/10.1103/physrevresearch.2.043440} {\bibfield  {journal} {\bibinfo  {journal} {Physical Review Research}\ }\textbf {\bibinfo {volume} {2}},\ \bibinfo {pages} {043440} (\bibinfo {year} {2020})}\BibitemShut {NoStop}%
\bibitem [{\citenamefont {Eisenlohr}\ and\ \citenamefont {Vojta}(2021)}]{Eisenlohr2021}%
  \BibitemOpen
  \bibfield  {author} {\bibinfo {author} {\bibfnamefont {H.}~\bibnamefont {Eisenlohr}}\ and\ \bibinfo {author} {\bibfnamefont {M.}~\bibnamefont {Vojta}},\ }\bibfield  {title} {\bibinfo {title} {Limits to magnetic quantum criticality from nuclear spins},\ }\href {https://doi.org/10.1103/physrevb.103.064405} {\bibfield  {journal} {\bibinfo  {journal} {Physical Review B}\ }\textbf {\bibinfo {volume} {103}},\ \bibinfo {pages} {064405} (\bibinfo {year} {2021})}\BibitemShut {NoStop}%
\bibitem [{\citenamefont {Bitko}\ \emph {et~al.}(1996)\citenamefont {Bitko}, \citenamefont {Rosenbaum},\ and\ \citenamefont {Aeppli}}]{Bitko1996}%
  \BibitemOpen
  \bibfield  {author} {\bibinfo {author} {\bibfnamefont {D.}~\bibnamefont {Bitko}}, \bibinfo {author} {\bibfnamefont {T.~F.}\ \bibnamefont {Rosenbaum}},\ and\ \bibinfo {author} {\bibfnamefont {G.}~\bibnamefont {Aeppli}},\ }\bibfield  {title} {\bibinfo {title} {Quantum critical behavior for a model magnet},\ }\href {https://doi.org/10.1103/physrevlett.77.940} {\bibfield  {journal} {\bibinfo  {journal} {Physical Review Letters}\ }\textbf {\bibinfo {volume} {77}},\ \bibinfo {pages} {940} (\bibinfo {year} {1996})}\BibitemShut {NoStop}%
\bibitem [{\citenamefont {Libersky}\ \emph {et~al.}(2021)\citenamefont {Libersky}, \citenamefont {McKenzie}, \citenamefont {Silevitch}, \citenamefont {Stamp},\ and\ \citenamefont {Rosenbaum}}]{Libersky2021}%
  \BibitemOpen
  \bibfield  {author} {\bibinfo {author} {\bibfnamefont {M.}~\bibnamefont {Libersky}}, \bibinfo {author} {\bibfnamefont {R.~D.}\ \bibnamefont {McKenzie}}, \bibinfo {author} {\bibfnamefont {D.~M.}\ \bibnamefont {Silevitch}}, \bibinfo {author} {\bibfnamefont {P.~C.~E.}\ \bibnamefont {Stamp}},\ and\ \bibinfo {author} {\bibfnamefont {T.~F.}\ \bibnamefont {Rosenbaum}},\ }\bibfield  {title} {\bibinfo {title} {Direct observation of collective electronuclear modes about a quantum critical point},\ }\href {https://doi.org/10.1103/physrevlett.127.207202} {\bibfield  {journal} {\bibinfo  {journal} {Physical Review Letters}\ }\textbf {\bibinfo {volume} {127}},\ \bibinfo {pages} {207202} (\bibinfo {year} {2021})}\BibitemShut {NoStop}%
\bibitem [{\citenamefont {Banda}\ \emph {et~al.}(2023)\citenamefont {Banda}, \citenamefont {Hafner}, \citenamefont {Landaeta}, \citenamefont {Hassinger}, \citenamefont {Mitsumoto}, \citenamefont {Giovannini}, \citenamefont {Sereni}, \citenamefont {Geibel},\ and\ \citenamefont {Brando}}]{banda2023}%
  \BibitemOpen
  \bibfield  {author} {\bibinfo {author} {\bibfnamefont {J.}~\bibnamefont {Banda}}, \bibinfo {author} {\bibfnamefont {D.}~\bibnamefont {Hafner}}, \bibinfo {author} {\bibfnamefont {J.~F.}\ \bibnamefont {Landaeta}}, \bibinfo {author} {\bibfnamefont {E.}~\bibnamefont {Hassinger}}, \bibinfo {author} {\bibfnamefont {K.}~\bibnamefont {Mitsumoto}}, \bibinfo {author} {\bibfnamefont {M.}~\bibnamefont {Giovannini}}, \bibinfo {author} {\bibfnamefont {J.~G.}\ \bibnamefont {Sereni}}, \bibinfo {author} {\bibfnamefont {C.}~\bibnamefont {Geibel}},\ and\ \bibinfo {author} {\bibfnamefont {M.}~\bibnamefont {Brando}},\ }\href@noop {} {\bibinfo {title} {Electronuclear quantum criticality}} (\bibinfo {year} {2023}),\ \Eprint {https://arxiv.org/abs/arXiv:2308.15294} {arXiv:2308.15294} \BibitemShut {NoStop}%
\bibitem [{\citenamefont {Bangma}\ \emph {et~al.}(2023)\citenamefont {Bangma}, \citenamefont {Levitin}, \citenamefont {Lucas}, \citenamefont {Casey}, \citenamefont {Nyeki}, \citenamefont {Broeders}, \citenamefont {Sutton}, \citenamefont {Andraka}, \citenamefont {Julian}, \citenamefont {Saunders},\ and\ \citenamefont {McCollam}}]{Bangma2023}%
  \BibitemOpen
  \bibfield  {author} {\bibinfo {author} {\bibfnamefont {F.}~\bibnamefont {Bangma}}, \bibinfo {author} {\bibfnamefont {L.}~\bibnamefont {Levitin}}, \bibinfo {author} {\bibfnamefont {M.}~\bibnamefont {Lucas}}, \bibinfo {author} {\bibfnamefont {A.}~\bibnamefont {Casey}}, \bibinfo {author} {\bibfnamefont {J.}~\bibnamefont {Nyeki}}, \bibinfo {author} {\bibfnamefont {I.}~\bibnamefont {Broeders}}, \bibinfo {author} {\bibfnamefont {A.}~\bibnamefont {Sutton}}, \bibinfo {author} {\bibfnamefont {B.}~\bibnamefont {Andraka}}, \bibinfo {author} {\bibfnamefont {S.}~\bibnamefont {Julian}}, \bibinfo {author} {\bibfnamefont {J.}~\bibnamefont {Saunders}},\ and\ \bibinfo {author} {\bibfnamefont {A.}~\bibnamefont {McCollam}},\ }\href {https://doi.org/10.48550/ARXIV.2305.17088} {\bibinfo {title} {Diverse influences of hyperfine interactions on strongly correlated electron states}} (\bibinfo {year} {2023}),\ \Eprint {https://arxiv.org/abs/arXiv:2305.17088} {arXiv:2305.17088} \BibitemShut {NoStop}%
\bibitem [{\citenamefont {Andres}\ and\ \citenamefont {Lounasmaa}(1982)}]{Andres1982}%
  \BibitemOpen
  \bibfield  {author} {\bibinfo {author} {\bibfnamefont {K.}~\bibnamefont {Andres}}\ and\ \bibinfo {author} {\bibfnamefont {O.~V.}\ \bibnamefont {Lounasmaa}},\ }\bibfield  {title} {\bibinfo {title} {Chapter 4: Recent progress in nuclear cooling},\ }in\ \href {https://doi.org/10.1016/s0079-6417(08)60007-4} {\emph {\bibinfo {booktitle} {Progress in Low Temperature Physics}}}\ (\bibinfo  {publisher} {Elsevier},\ \bibinfo {year} {1982})\ pp.\ \bibinfo {pages} {221--287}\BibitemShut {NoStop}%
\end{thebibliography}

%

\end{document}